\documentclass[useAMS,usenatbib]{mn2e}
\usepackage{graphicx,natbib,amssymb,amsmath,dcolumn,color,placeins}
\newcommand{\figref}[1]{figure~\ref{#1}}
\newcommand{\figsref}[1]{figures~\ref{#1}}
\newcommand{\figsand}[2]{figures~\ref{#1} and \ref{#2}}
\newcommand{\Figref}[1]{Figure~\ref{#1}}
\newcommand{\Figsref}[1]{Figures~\ref{#1}}
\newcommand{\secref}[1]{section~\ref{#1}}
\newcommand{\secsand}[2]{sections~\ref{#1} and \ref{#2}}

\newcommand{\apref}[1]{appendix~\ref{#1}}

\renewcommand{\eqref}[1]{equation~(\ref{#1})}
\newcommand{\Eqsref}[1]{Equations~(\ref{#1})}

\newcommand{\exref}[1]{(\ref{#1})}

\newcommand{\bea}{\begin{eqnarray}}
\newcommand{\eea}{\end{eqnarray}}
\newcommand{\beq}{\begin{equation}}
\newcommand{\eeq}{\end{equation}}
\newcommand{\lt}{\left}
\newcommand{\rt}{\right}

\renewcommand{\la}{\langle}
\newcommand{\ra}{\rangle}
\newcommand{\dd}{\partial}

\newcommand{\eps}{\varepsilon}
\newcommand{\rmd}{\rmn{d}}
\renewcommand{\Re}{\rmn{Re}}
\newcommand{\Reff}{\Re_\mathrm{eff}}
\newcommand{\Ma}{\rmn{Ma}}

\newcommand{\lvisc}{l_{\mathrm{visc}}}
\newcommand{\betac}{\beta_{\mathrm{c}}}
\newcommand{\opi}{\omega_{\mathrm{p}i}}
\newcommand{\vth}{v_{\mathrm{th}}}

\newcommand{\nuii}{\nu_{ii}} 
\newcommand{\nueff}{\nu_{\rm eff}} 
\newcommand{\mueff}{\mu_{\rm eff}} 
\newcommand{\tauc}{\tau_{\rm c}} 

\newcommand{\tsat}{t_{\rm sat}}

\newcommand{\pperp}{p_\perp}
\newcommand{\ppar}{p_\parallel}

\newcommand{\tS}{\tilde S}

\newcommand{\vB}{\bmath{B}}
\newcommand{\dvB}{\delta\vB}
\newcommand{\dB}{\delta B}
\newcommand{\dBperp}{\dB_\perp}
\newcommand{\dvBperp}{\dvB_\perp}
\newcommand{\dBpar}{\dB_\parallel}
\newcommand{\vb}{\bmath{b}}
\newcommand{\dvb}{\delta\vb}
\newcommand{\bsq}{\la|\dvb|^2\ra}
\newcommand{\vE}{\bmath{E}}

\newcommand{\vA}{v_{\rm A}}
\newcommand{\vu}{\bmath{u}}
\newcommand{\vj}{\bmath{j}}

\newcommand{\vv}{\bmath{v}}
\newcommand{\vperp}{v_\perp}

\newcommand{\vpar}{v_\parallel}

\newcommand{\vx}{\hat{\bmath{x}}}
\newcommand{\vy}{\hat{\bmath{y}}}

\newcommand{\vdel}{\bmath{\nabla}}

\newcommand{\lpar}{l_\parallel}
\newcommand{\lpare}{l_\parallel^\mathrm{(edge)}}
\newcommand{\lparm}{l_\parallel^\mathrm{(mirror)}}
\newcommand{\lperp}{l_\perp}
\newcommand{\kperp}{k_\perp}
\newcommand{\kpar}{k_\parallel}

\title[Pressure-anisotropy microturbulence and magnetogenesis]{Pressure-anisotropy-driven microturbulence 
and magnetic-field evolution in shearing, collisionless plasma}
\author[S.~Melville, A.~A.~Schekochihin and M.~W.~Kunz]{
Scott~Melville,$^{1,2,3,4}$\thanks{E-mail: scott.melville@oxon.org}
Alexander~A.~Schekochihin$^{1,5}$\thanks{E-mail: alex.schekochihin@physics.ox.ac.uk}
and
Matthew~W.~Kunz$^{6,7}$\thanks{E-mail: mkunz@princeton.edu}\\
$^1$The Rudolf Peierls Centre for Theoretical Physics, University of Oxford, Oxford OX1 3NP, UK\\
$^2$The Queen's College, Oxford OX1 4AW, UK\\
$^3$Wolfgang Pauli Institute, University of Vienna, 1090 Vienna, Austria \\
$^4$Harvard University, Cambridge, Massachusetts 02138, USA\\ 
$^5$Merton College, Oxford OX1 4JD, UK\\
$^6$Department of Astrophysical Sciences, Princeton University, Princeton, New Jersey 08544, USA\\
$^7$Princeton Plasma Physics Laboratory, Princeton, New Jersey 08543, USA}

\begin{document}

\date{\today}

\pagerange{\pageref{firstpage}--\pageref{lastpage}} \pubyear{2014}

\maketitle

\label{firstpage}

\begin{abstract}

The nonlinear state of a high-beta collisionless plasma is investigated 
in which an externally imposed linear shear amplifies or diminishes 
a uniform mean magnetic field, driving pressure anisotropies 
and, therefore, firehose or mirror instabilities. 
The evolution of the resulting microscale turbulence is considered 
when the external shear {\em changes}, mimicking the local behaviour of 
a macroscopic turbulent plasma flow, {\em viz.}, the shear is either switched off 
or reversed after one shear time, so that a new macroscale configuration 
is superimposed on the microscale state left behind by the previous one. 
It is shown that there is a threshold value of plasma beta: when $\beta\ll\Omega/S$
(ion cyclotron frequency/shear rate), the emergence of firehose or mirror 
fluctuations when they are driven unstable by shear and their disappearance 
when the shear is removed or reversed are quasi-instantaneous compared 
to the macroscopic (shear) timescale. This is because the free-decay time scale 
of these fluctuations is $\sim\beta/\Omega \ll S^{-1}$, a result that arises 
from the free decay of both types of fluctuations turning out to be constrained by 
the same marginal-stability thresholds as their growth and saturation 
in the driven regime. 
In contrast, when $\beta\gtrsim\Omega/S$ (``ultra-high'' beta), 
the old microscale state can only be removed on the macroscopic (shear) timescale. 
Furthermore, it is found that in this ultra-high-beta regime, 
when the firehose fluctuations are driven, they grow secularly to order-unity 
amplitudes (relative to the mean field), this growth compensating for the decay 
of the mean field, with pressure anisotropy pinned at marginal stability purely 
by the increase in the fluctuation energy, with no appreciable scattering 
of particles (which is unlike what happens at moderate $\beta$). 
When the shear reverses, the shearing away of this firehose turbulence 
compensates for the increase in the mean field and thus  
prevents growth of the pressure anisotropy, stopping the system from 
going mirror-unstable. Therefore, at ultra-high $\beta$, the system stays 
close to the firehose threshold, the mirror instability is almost 
completely suppressed, while the mean magnitude of the magnetic field barely 
changes at all. Implications of the properties of both ultra-high- and 
moderate-beta regimes for the operation of plasma dynamo and thus the origin 
of cosmic magnetism are discussed, suggesting that collisionless effects 
are broadly beneficial to fast magnetic-field generation. 
\end{abstract}


\section{Introduction}
\label{sec:intro}

In recent years, it has become increasingly clear that the dynamics of a high-beta, 
collisionless (or weakly collisional) plasma are both a poorly understood and very 
interesting subject. Astrophysically, this type of plasma 
represents most of the luminous matter in the Universe as it fills the intracluster 
medium (ICM) of the clusters of galaxies. 
The ICM is ``high-beta'' in the sense that 
the ratio of thermal to magnetic energy densities is large: 
\beq
\beta = \frac{8\pi p}{B^2} \gg 1, 
\eeq 
where $p$ is pressure and $B$ is magnetic-field strength (in galaxy-cluster cores, 
the typical values are $B\sim10^{-6}$~G and $\beta\sim10^2$; see, e.g., 
\citealt{Ensslin06} or \citealt{Rosin11} for a good set of fiducial ICM parameters). 
The reason for the large value of $\beta$ is that the 
magnetic-energy density tends to be comparable to the energy density of the 
turbulent plasma motions and the latter are typically subsonic, so 
$\beta \sim 1/\Ma^2$, where $\Ma\sim 0.1$ is the Mach number
\citep[e.g.,][]{Fujita05,Rebusco06,Sanders13,Zhuravleva14}. 
The ICM is ``weakly collisional'' in the sense that collisions 
are much less frequent than the Larmor gyration, 
\beq
\frac{\nuii}{\Omega} \ll 1,
\eeq
where $\nuii$ is the ion-ion collision frequency (ICM is a hydrogen plasma) 
and $\Omega$ is the ion Larmor frequency (assuming ion number density 
$n_i\sim10^{2}$~cm$^{-3}$ and temperature $T\sim10^7$~K, the ratio of these 
frequencies is small provided $\beta \ll 10^{24}$, or $B\gg10^{-17}$~G). 

The reason the dynamics 
of such plasmas are unlike those of an MHD fluid is that weak collisionality 
permits approximate conservation of particle adiabatic invariants and so 
when frozen-in magnetic field is locally increased or decreased by the 
plasma motions, perpendicular ($\pperp$) and parallel ($\ppar$) pressures evolve 
differently \citep{CGL,Kulsrud83}, giving rise to local pressure anisotropies. Pressure-anisotropic 
plasma is unstable when the relative pressure anisotropy is large enough compared to 
$1/\beta$ (otherwise Lorentz forces stabilise the plasma), 
namely,
\begin{align}
\label{eq:fh}
&\Delta \equiv \frac{\pperp-\ppar}{p} < -\frac{2}{\beta} 
&\Rightarrow\qquad& \rmn{firehose~instability},\\
&\Delta \equiv \frac{\pperp-\ppar}{p} > \frac{1}{\beta} 
&\Rightarrow\qquad& \rmn{mirror~instability}
\label{eq:mirr}
\end{align}
(the mirror threshold given here is approximate; see \citealt{Hellinger07}).
These instability conditions are 
easily achievable in the ICM (see \citealt{Sch05} and references therein). 
The instabilities are very fast in time and microscopic in space compared 
to any conceivable macroscopic dynamics because they happen essentially on ion Larmor scales. 
It is reasonable to expect that their 
effect will be to keep the plasma locally (at most) marginal with respect 
to the conditions 
\exref{eq:fh} and \exref{eq:mirr}; solar-wind measurements lend strong 
support to this expectation \citep{Hellinger06,Bale09}. 
In order to make that happen, the firehose and mirror fluctuations 
must impose an upper bound on the effective rate of change of the magnetic 
field (which gives rise to the pressure anisotropies) and/or increase the effective 
collisionality of the plasma (which relaxes the anisotropies)---either scenario can 
lead to macroscopic dynamics dramatically different from the conventional 
MHD evolution.

This topic was discussed at length (if speculatively) by \citet{Mogavero14}, 
with particular focus on the question of how, under pressure-anisotropic conditions, 
the magnetic field might be able to grow from a weak primordial seed (believed to be anywhere 
between $10^{-21}$ and $10^{-9}$~G, corresponding to $\beta\sim 10^{32}-10^{8}$; see 
review by \citealt{Durrer13}) to its observed value ($B\sim10^6$~G, $\beta\sim10^2$), 
even as the rate at which it is allowed to change 
locally is constrained by the condition 
\beq
\frac{1}{\nueff}\frac{1}{B}\frac{\rmd B}{\rmd t} \sim 
\Delta \equiv\frac{\pperp-\ppar}{p} \in \lt[-\frac{2}{\beta},\frac{1}{\beta}\rt],  
\label{eq:marg}
\eeq 
where $\nueff$ is either the Coulomb collision frequency or some effective collisionality
determined by particle scattering (if any) off microscopic firehose or mirror 
fluctuations (see further discussion in \secref{sec:mag}). 
This is the question of {\em plasma dynamo}---magnetogenesis in a weakly collisional 
turbulent plasma. Besides being of astrophysical (origin of observed fields) 
and fundamental (what happens?) physical interest, it is likely soon to be 
attacked not just theoretically, but also experimentally, with the advent of laboratory experiments 
aiming to reproduce the plasma dynamo process \citep{Spence09,Meinecke14,Meinecke15,Plihon15,Forest15}.  
The first demonstration of this process in a 3D kinetic numerical simulation 
by \citet{Rincon16} has indeed confirmed the ubiquitous appearance of mirror and firehose 
fluctuations, although a detailed investigation of the full multiscale 
problem remains computationally too intensive to be affordable.  

There are, obviously, numerous other problems and situations, unrelated to 
magnetogenesis or to the ICM, in which it matters how a plasma system behaves when it has to effect 
a local change of the magnetic field. Indeed, in most problems of astrophysical fluid dynamics, 
the magnetic field and the fluid (plasma) flow are linked, the former being frozen 
into the latter, and so very little can happen without magnetic field changing 
under the action of time-dependent velocity gradients. 

The type of modelling of the pressure-anisotropy effect on dynamics advocated 
by \citet{Mogavero14} (see also \citealt{Samsonov01,Sch06,Lyutikov07,Kunz11}), 
as well as in various numerical implementations of 
effective closures ensuring that $\Delta$ does not stray into the unstable 
regimes \exref{eq:fh} and \exref{eq:mirr} 
\citep{Sharma06,Sharma07,Samsonov07,Chandran11,Meng12,Kunz12,Lima14},
rests on the assumption that, as the magnetic field is locally 
increased (decreased), the system will stray into mirror (firehose) 
unstable regimes, instabilities will instantly produce corresponding fluctuations, 
those will saturate and adjust the instability thresholds (or the system's 
position with respect to them). This adjustment 
is assumed to happen very quickly compared to the time scale on which the system 
might switch from one instability threshold to the other. Thus, each time such 
a switch occurs, no memory of previous microscale fluctuations is assumed 
to be preserved---or, at least, to matter to the macroscopic evolution. 

This is certainly a reasonable expectation with regard to 
triggering the microscale fluctuations: they happen on Larmor timescales, which 
are always short compared to the time scale on which local macroscale magnetic field 
changes, because the latter is set by the local shear, rate of strain 
or divergence associated with slow, macroscopic motions (we will henceforth 
refer simply to local shear, but the effect of other types of field-changing 
motions is analogous). In a turbulent plasma, 
the local shear time is also the time scale over which 
the corresponding plasma motion is correlated and so over which 
a particular local episode of increasing (decreasing) magnetic field will persist. 
When that ends and the field starts decreasing (increasing) under the influence 
of a new local shear of, in general, opposite sign and different magnitude, 
the system is not in fact 
in a ``clean state''---there is a sea of residual microscale fluctuations that were 
keeping the plasma locally marginal with respect to the old shear. 
Do they simply decay away to free the system to develop another set of 
fluctuations suitable for the new shear? Can one, therefore, model 
macroscopic dynamics in the way that has been emerging in the literature
(see above)? 
This turns out to be a nontrivial question, the answer to which depends 
on the magnitude of $\beta$ and has interesting implications for 
the expected scenario of magnetic-field amplification in astrophysical 
environments such as the ICM. 
It is this question that will preoccupy us in this paper 
and we will show, in particular, that at ultra-high $\beta$ (very weak fields), 
plasma develops a persistent multiscale state, with the 
dynamics of the macroscale fields, pressure anisotropy and microscale 
(specifically, firehose) fluctuations inextricably linked. 

The plan of further proceedings is as follows. 
In \secref{sec:decay}, we will give a brief review of the behaviour of firehose 
and mirror fluctuations in a shearing plasma and then show how these 
fluctuations decay when the shear is switched off. The key idea will be that 
the marginality condition \exref{eq:marg} constrains the free decay of microscale 
turbulence just as it does the rate of change of the magnetic field under shearing. 
We will also find a new type of driven firehose regime at ultra-high $\beta$, 
in which  fluctuations grow for at least a shear time without scattering particles 
(\secref{sec:fh_sat}). 
In \secref{sec:shear}, we will investigate what happens when a shear reverses. 
The most consequential conclusion will be that, at high enough $\beta$, 
switching from decreasing 
to increasing magnetic field does not cause the system to cross the mirror 
threshold because of the mitigating effect that residual decaying firehose  
fluctuations have on the pressure anisotropy (\secref{sec:fh_to_mr_num}). 
In \secref{sec:conc}, we will give an itemised summary of our findings 
(\secref{sec:sum}) and discuss their implications for the theory of 
the origin of cosmic magnetism (\secref{sec:mag}). 
The conclusions with regard to the latter are broadly optimistic: 
generally speaking, microphysics does not appear to oppose field growth but 
is reluctant to allow the field to decay, so the plasma dynamo may
be an easier proposition than the MHD one.\footnote{The idea that plasma dynamo 
accelerates as the field grows \citep{Sch06,Mogavero14} appears to have just received 
some numerical support \citep{Rincon16}.} 

\begin{figure}
\includegraphics[width=0.45\textwidth]{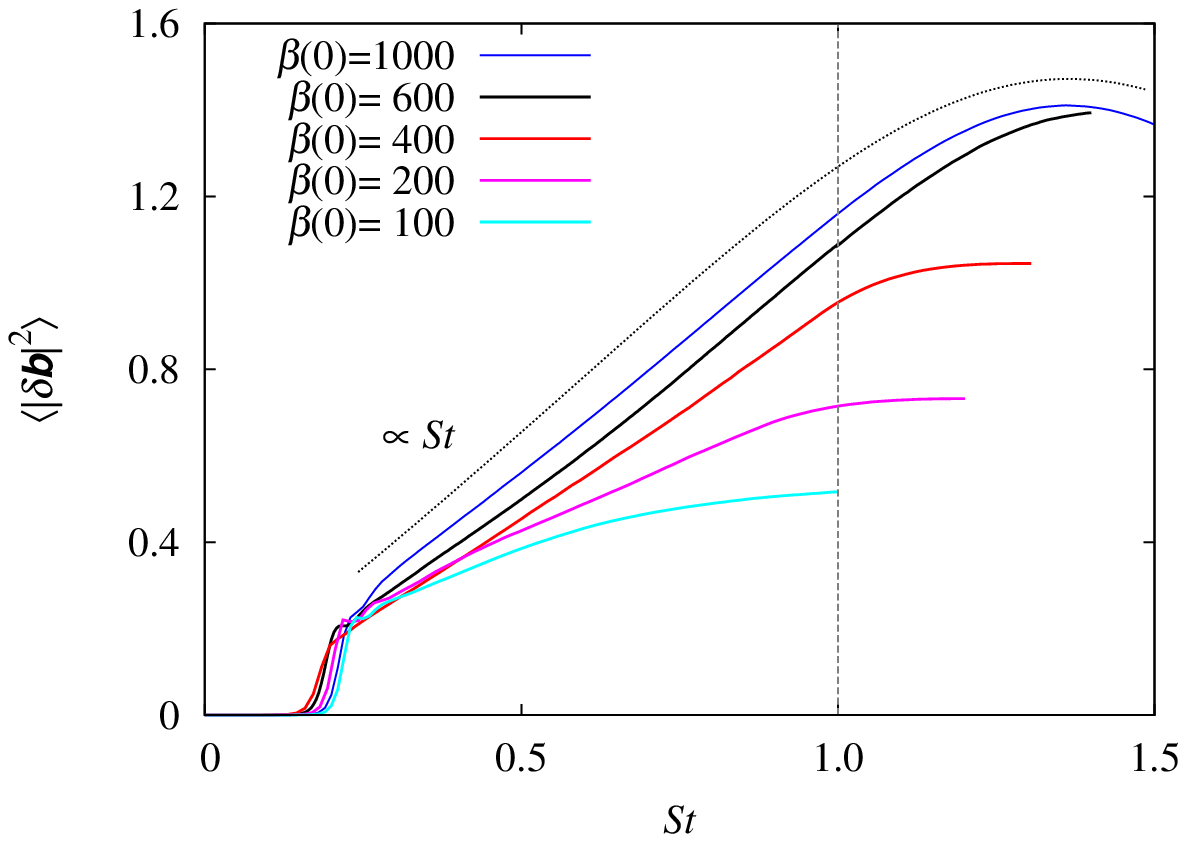}
\caption{Time evolution of the firehose fluctuation level, $\bsq$, 
for runs with $S=2\times10^{-3}\Omega$ and different initial $\beta$. 
The dotted line represents the analytical prediction for secular growth (second line of \eqref{eq:fh_sec}, 
where $\vB_0(0)$ is in fact taken at $St=0.25$, the beginning of the secular-growth stage).
Here and in all other plots, $\bsq=\la|\dvBperp|^2\ra/B_0^2$, including the time dependence 
of the mean field, $B_0=B_0(t)$ (see \eqref{eq:Bo_t}).}
\label{fig:fh_growth}
\end{figure}

\section{Growth, saturation and decay of firehose and mirror fluctuations}
\label{sec:decay}

\subsection{Growth and saturation of firehose fluctuations}
\label{sec:fh_sat}

\subsubsection{Theoretical expectations}

Let us start by positing that the mean (ion) pressure anisotropy 
evolves according to 
\beq
\frac{\rmd\Delta}{\rmd t} = 3\lt\la\frac{\rmd\ln B}{\rmd t}\rt\ra - 3\nueff\Delta.
\label{eq:Delta}
\eeq
This follows essentially from CGL equations, with account taken of the isotropising 
effect of collisions \citep[see][]{Sch10,Rosin11}. 
We have assumed that $\pperp-\ppar\ll p$, that density and temperature 
do not change with time, that there are no heat fluxes and that 
$\nueff$ is the larger of the Coulomb collision frequency and the 
anomalous scattering rate of particles off microscale fluctuations (if any).  
The angle brackets mean averaging over space (i.e., over the microscales) 
and we assume that the total magnetic field consists of a slowly 
decreasing mean field $\vB_0=\la\vB\ra$ and growing firehose 
fluctuations $\dvBperp$ (perpendicular to the mean field) excited by 
the resulting pressure anisotropy. Then, assuming small fluctuations, 
\beq
\la\ln B\ra \approx \ln B_0 + \frac{1}{2}\,\bsq, 
\quad
\dvb \equiv \frac{\dvBperp}{B_0},
\label{eq:dB_fh}
\eeq 
and \eqref{eq:Delta} becomes
\beq
\frac{1}{3}\frac{\rmd\Delta}{\rmd t} + \nueff\Delta 
= -\lt|\frac{\rmd\ln B_0}{\rmd t}\rt| + \frac{1}{2}\frac{\rmd\bsq}{\rmd t},
\label{eq:Delta_fh}
\eeq
where we have accentuated the fact that the rate of change of the mean field is negative. 
In a simple model system where the decrease of the mean field is achieved by persistent 
linear shear flow $\vu = -S x \vy$ \citep{Kunz14}, we have 
\beq
\frac{\rmd\ln B_0}{\rmd t} = -\frac{B_{0x}B_{0y}}{B_0^2}\,S \equiv - \tS(t), 
\label{eq:gamma0}
\eeq
which is approximately constant in time, $\tS\sim S$, at times $t\ll S^{-1}$. 
At longer times, the slow shearing away of the mean field,
\beq
B_0(t) = \sqrt{B_{0x}^2 + \lt[B_{0y}(0) - B_{0x} St\rt]^2}, 
\label{eq:Bo_t}
\eeq
must be taken into account:
\beq
\tS(t) = \frac{B_{0x}\lt[B_{0y}(0) - B_{0x} St\rt]}{B_{0x}^2 
+ \lt[B_{0y}(0) - B_{0x} St\rt]^2}\,S. 
\label{eq:gg_exact}
\eeq

\begin{figure*}
\begin{tabular}{cc}
\includegraphics[width=0.45\textwidth]{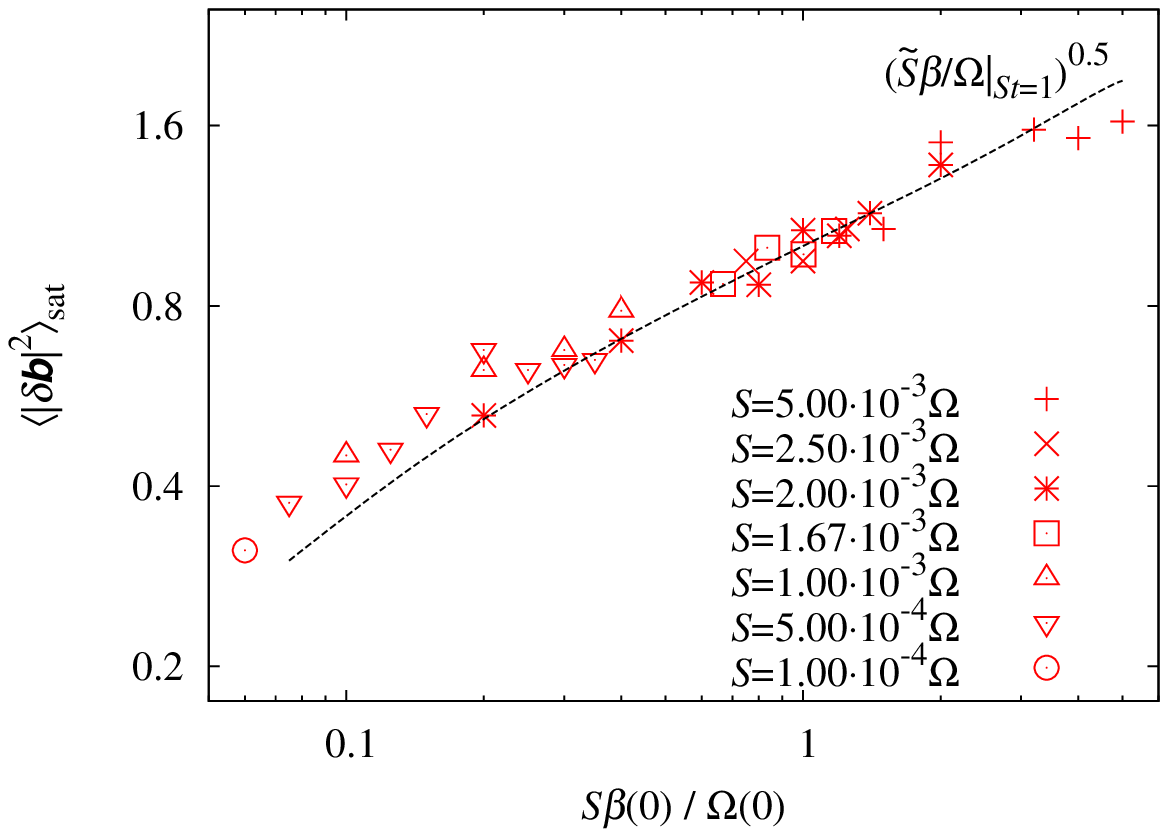}&
\includegraphics[width=0.45\textwidth]{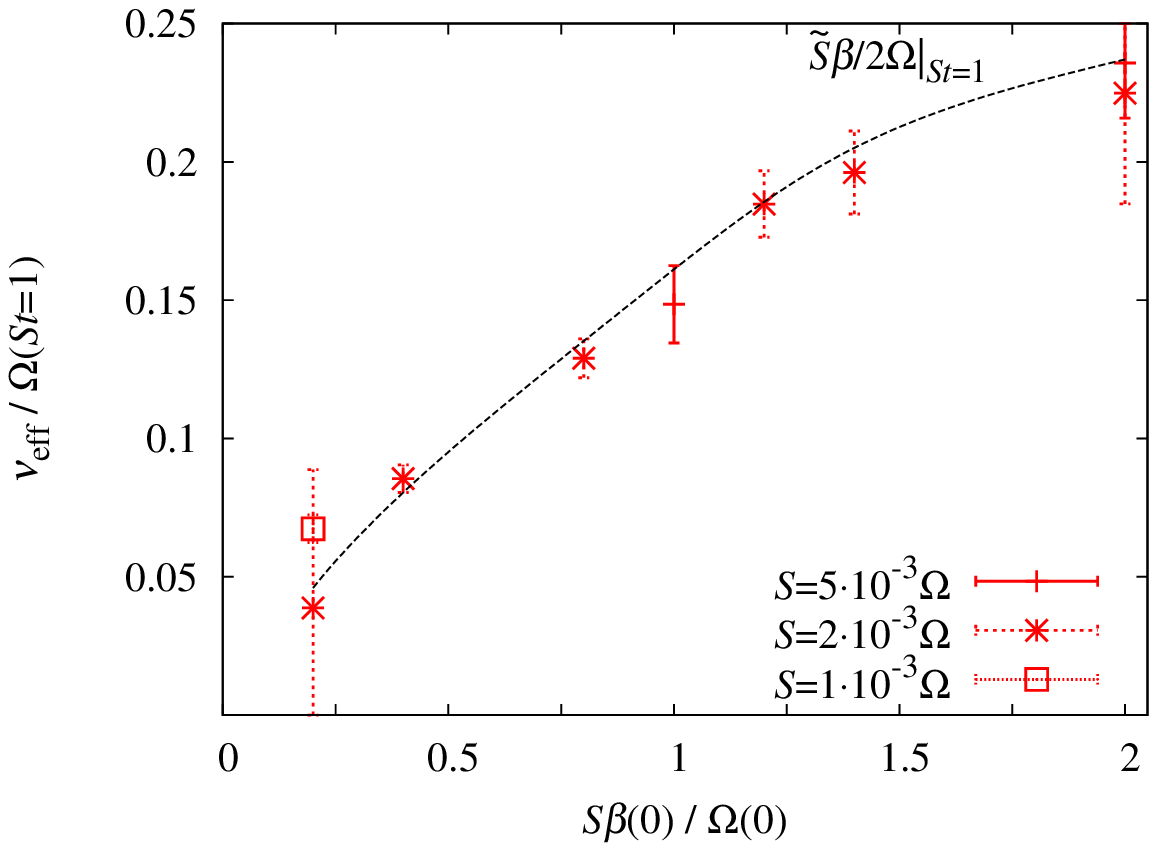}\\
(a)&(b)
\end{tabular}
\caption{Dependence of (a) the saturated level of driven firehose fluctuations $\bsq$ 
and (b) their effective scattering rate $\nueff/\Omega$ 
(calculated by the method described in \apref{ap:coll})
on shear $S$ and plasma beta $\beta$.
The dotted line in (a) is the empirical scaling \exref{eq:fh_sat}, 
including a fitting constant and using the value of $\beta$ at $St=1$: 
$\bsq\approx 0.77(\beta\tS/\Omega)^{1/2}$, where $\tS$ is defined in 
\eqref{eq:gamma0}. This scaling is not theoretically understood, 
but manifestly quite well satisfied except when $\beta \gtrsim \Omega/S$. 
The dotted line in (b) is the theoretical value \exref{eq:fh_nueff} required 
for the anomalous scattering to pin the anisotropy at marginal level 
(where it is indeed pinned; see \figref{fig:fh_decay}b).}
\label{fig:fh_sat}
\end{figure*}

Assuming further that the system (after an initial transient) will settle into a marginal state, 
\beq
\Delta \to -\frac{2}{\beta},
\label{eq:Delta_fh_marg}
\eeq
we find from \eqref{eq:Delta_fh} that in such a state, 
\beq
-\frac{2\nueff}{\beta}
= -\tS + \frac{1}{2}\frac{\rmd\bsq}{\rmd t} 
\label{eq:fh_marg}
\eeq
(we assume $\beta\gg1$ and so the $\rmd\Delta/\rmd t$ term is small). 
In a driven system where $B_0$ is continuously decreased,
the marginality \exref{eq:fh_marg} can be achieved in two ways. 
In the absence of anomalous scattering ($\nueff\ll S\beta$), 
the rate of growth of the total energy of the firehose fluctuations can cancel the 
rate of decrease of the mean field, leading to secularly growing 
``firehose turbulence'' \citep{Sch08,Rosin11}:
\begin{align}
\nonumber
\frac{1}{2}\bsq(t) &= \int^t_0\rmd t'\tS(t')\\
\nonumber
&=\ln\lt[1 - \frac{2B_{0y}(0)B_{0x}}{B_0^2(0)}\,St
 + \frac{B_{0x}^2}{B_0^2(0)}(St)^2\rt]^{-1/2}\\ 
&\sim St.
\label{eq:fh_sec}
\end{align} 
Alternatively, if the firehose fluctuations are capable of scattering particles, 
they can happily saturate provided the scattering rate satisfies 
\beq
\nueff = \frac{\tS\beta}{2} \sim S\beta. 
\label{eq:fh_nueff}
\eeq
Which of these effects takes precedence in reality? 

\begin{figure*}
\begin{tabular}{cc}
\includegraphics[width=0.45\textwidth]{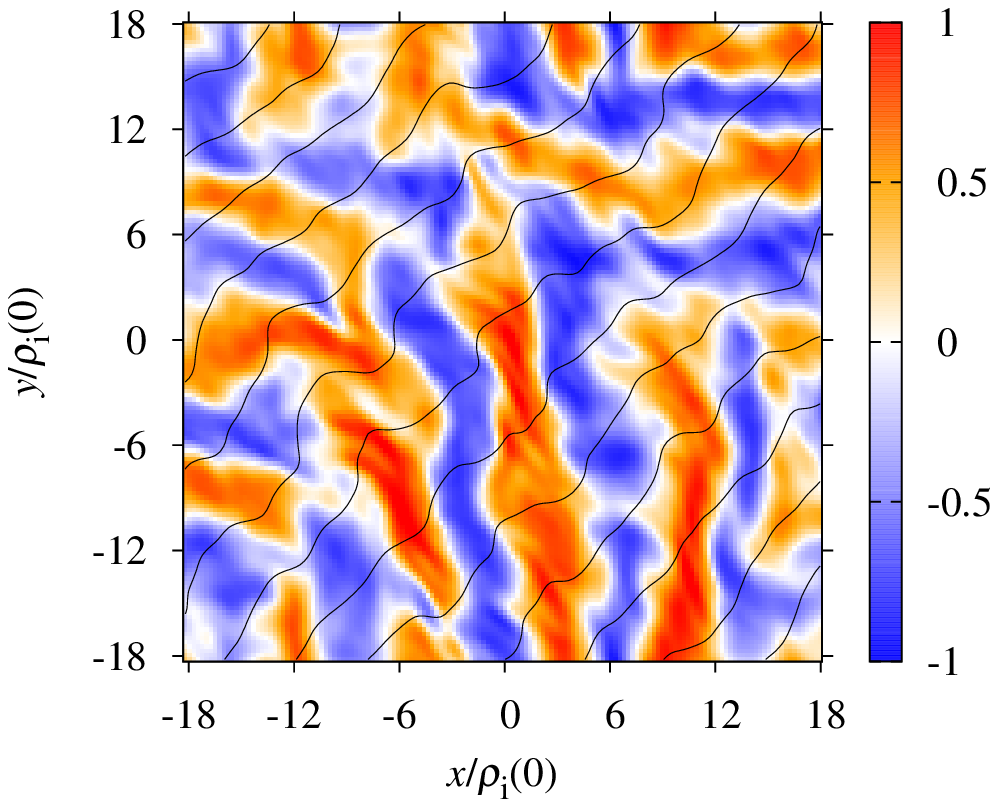}&
\includegraphics[width=0.45\textwidth]{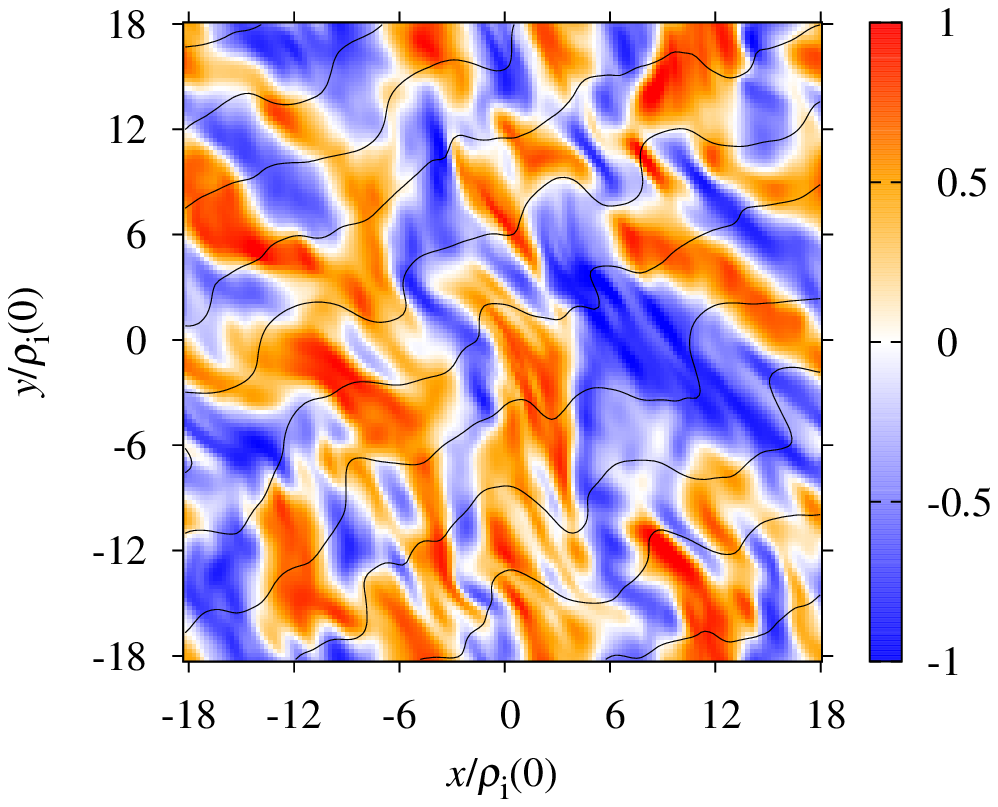}\\
(a) Growing firehose (no scattering), $\dB_z(x,y)$ at $St = 0.4$ & 
(b) Growing firehose (with scattering), $\dB_z(x,y)$ at $St = 1.0$\\\\
\includegraphics[width=0.45\textwidth]{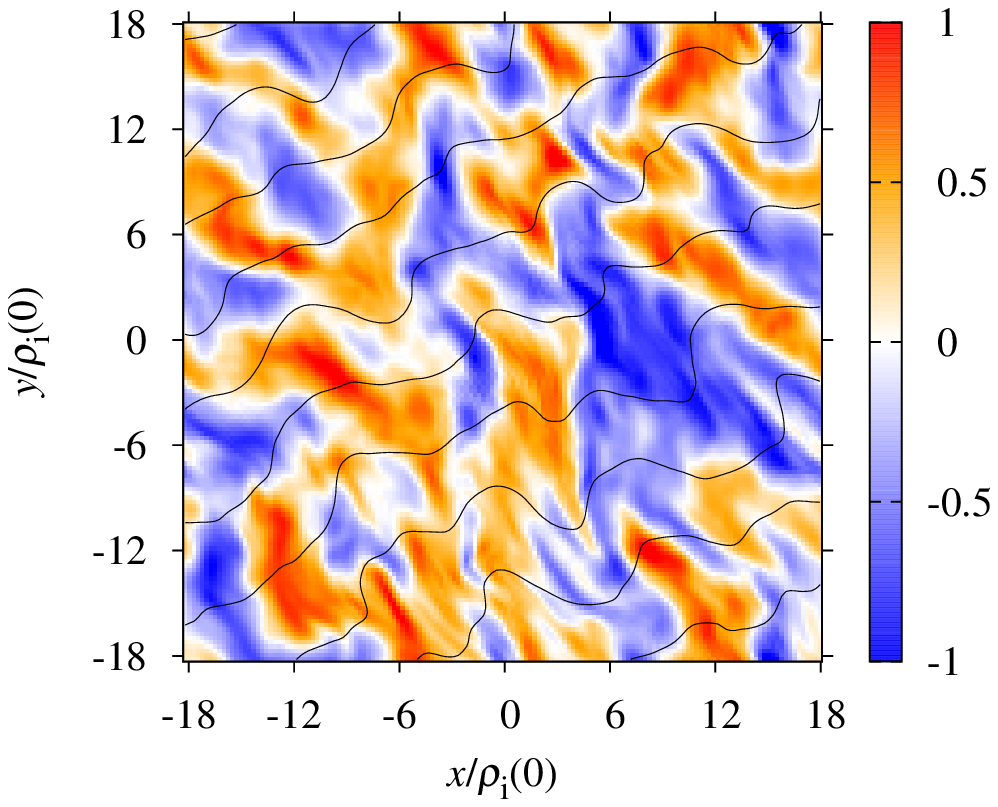}&
\includegraphics[width=0.45\textwidth]{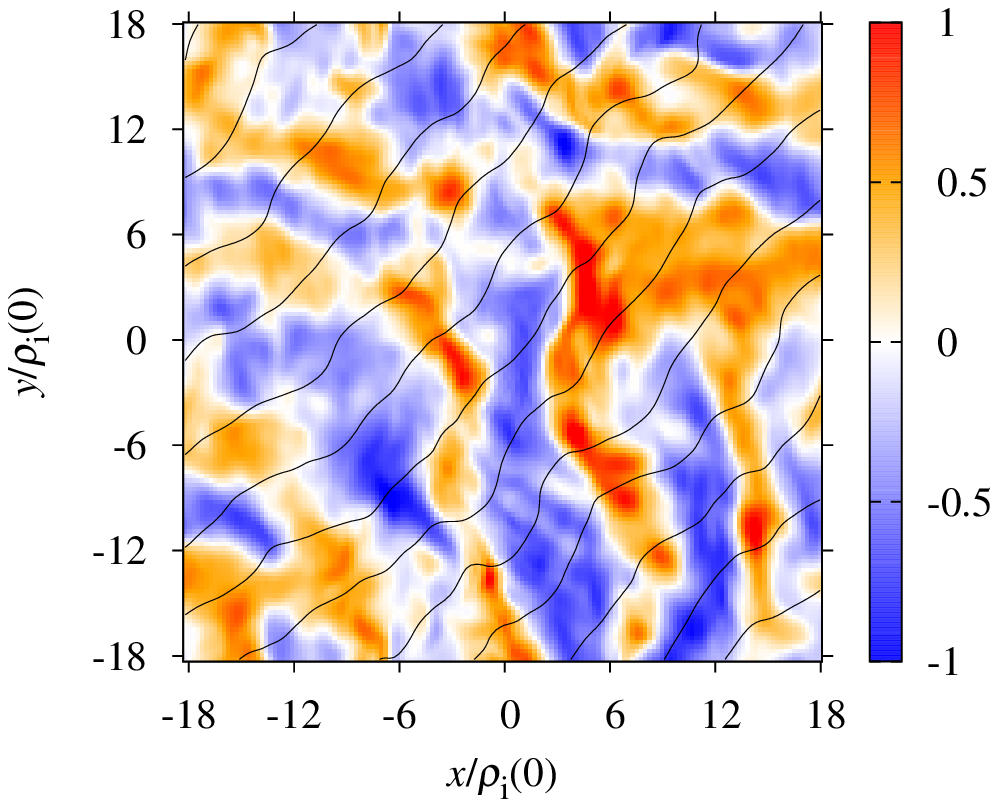}\\
(c) Decaying firehose, $\dB_z(x,y)$ at $St = 1.4$ & 
(d) From firehose to mirror, $\dB_z(x,y)$ at $St=1.4$\\\\
\includegraphics[width=0.45\textwidth]{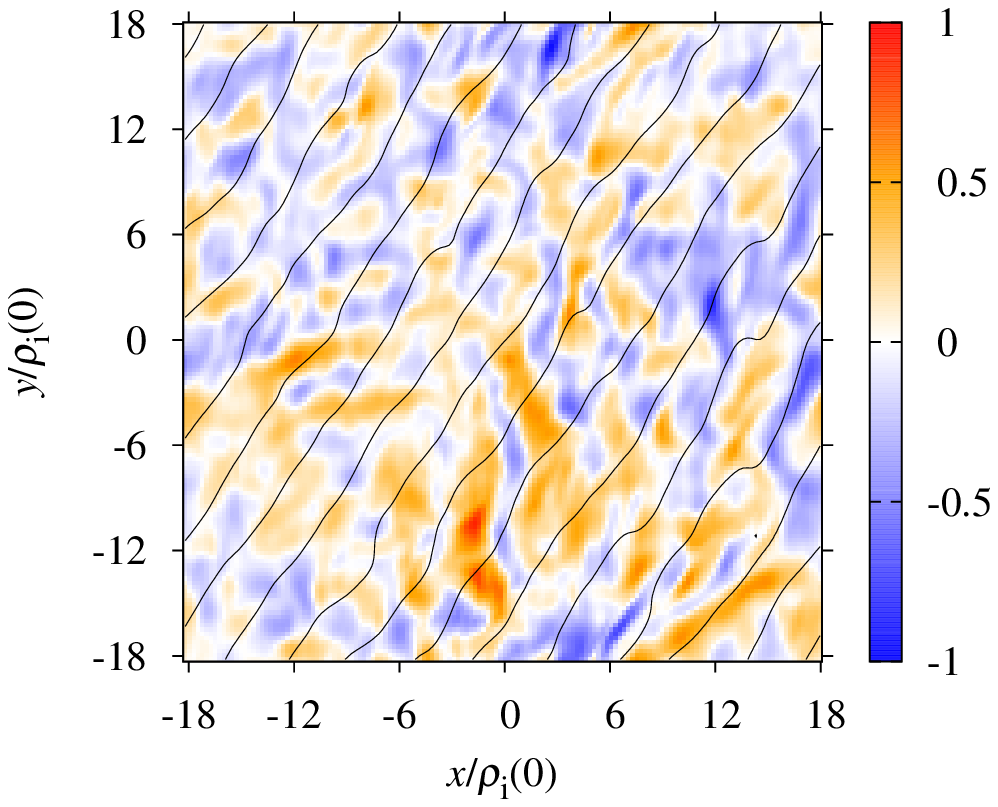}&
\includegraphics[width=0.45\textwidth]{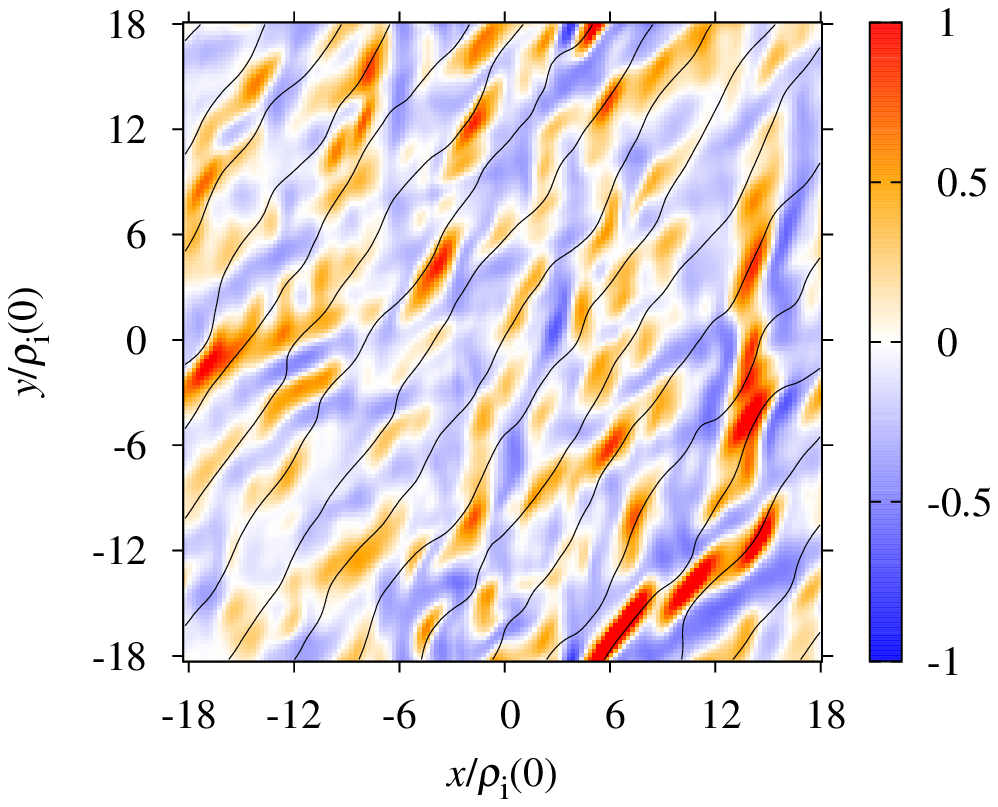}\\
(e) From firehose to mirror, $\dB_z(x,y)$ at $St=2.0$ & 
(f) From firehose to mirror, $\dBpar(x,y)$ at $St=2.0$
\end{tabular}
\caption{A series of snapshots of the fluctuating fields in simulations 
with $\beta(0)=600$ and $S=2\times10^{-3}\Omega$,  
starting in the firehose-unstable regime: (a) $\dB_z(x,y)$ (proxy 
for $\dvBperp$) at $St=0.4$ (growing firehose perturbations, no anomalous scattering; 
see \figref{fig:fh_decay}a and d); 
(b) same as (a) at $St=1.0$ (growing firehose perturbations, but now with 
substantial anomalous scattering; 
see \figref{fig:fh_decay}a and d);
(c) same as (b) at $St=1.4$ in a run where shear was switched off at $St=1$ 
(decaying firehose perturbations; see \figref{fig:fh_decay}a));
(d) same as (b) at $St=1.4$ in a run where shear was reversed at $St=1$
(firehose perturbations being sheared away; see \figref{fig:fh_to_mr}a);
(e) same as (d) at $St=2.0$ in the run with reversed shear
(firehose perturbations are nearly gone; see \figref{fig:fh_to_mr}a));
(f) $\dBpar(x,y)$ in the same run and at the same time as (e) 
(mirror perturbations starting to grow; see \figref{fig:fh_to_mr}a). 
The $x$ and $y$ coordinates are shown in the units of the ion Larmor
radius at the start of the simulation, $\rho_i(0)$. 
Black lines are magnetic-field lines.} 
\label{fig:fh_pics}
\end{figure*}

\begin{figure*}
\begin{tabular}{cc}
\includegraphics[width=0.5\textwidth]{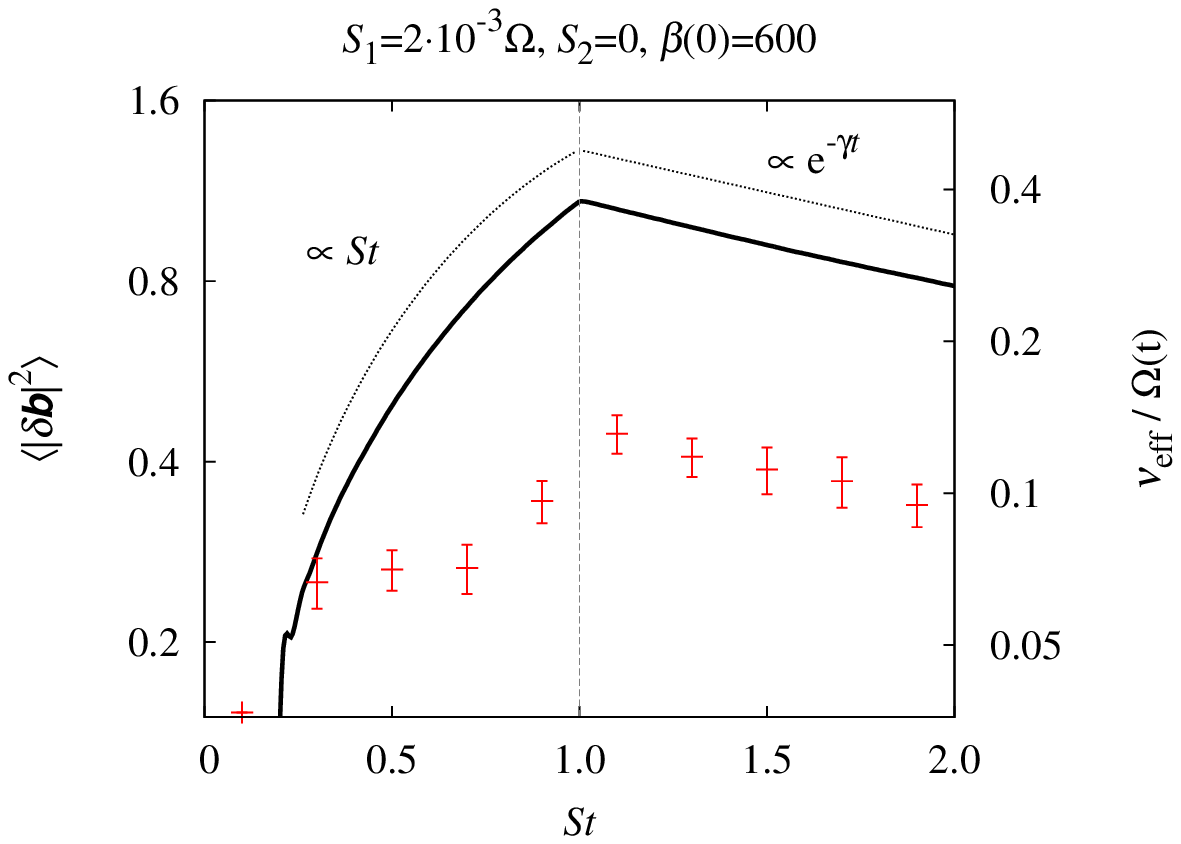}&
\includegraphics[width=0.5\textwidth]{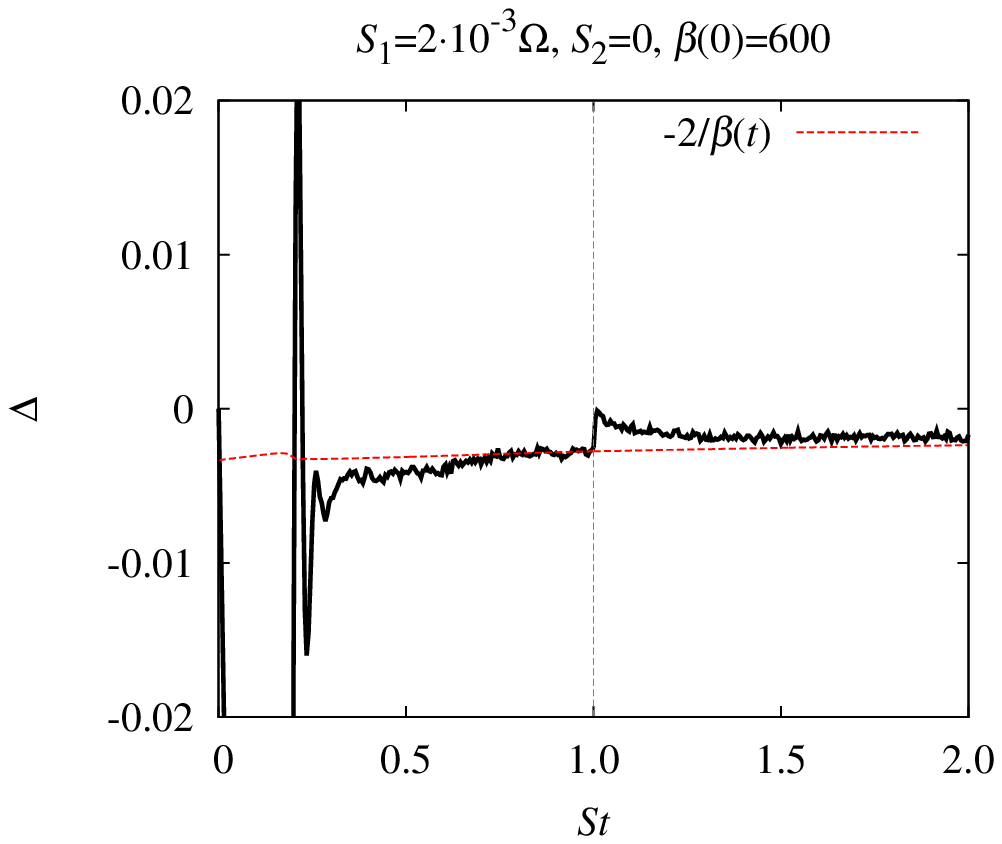}\\
(a)&(b)\\
\includegraphics[width=0.5\textwidth]{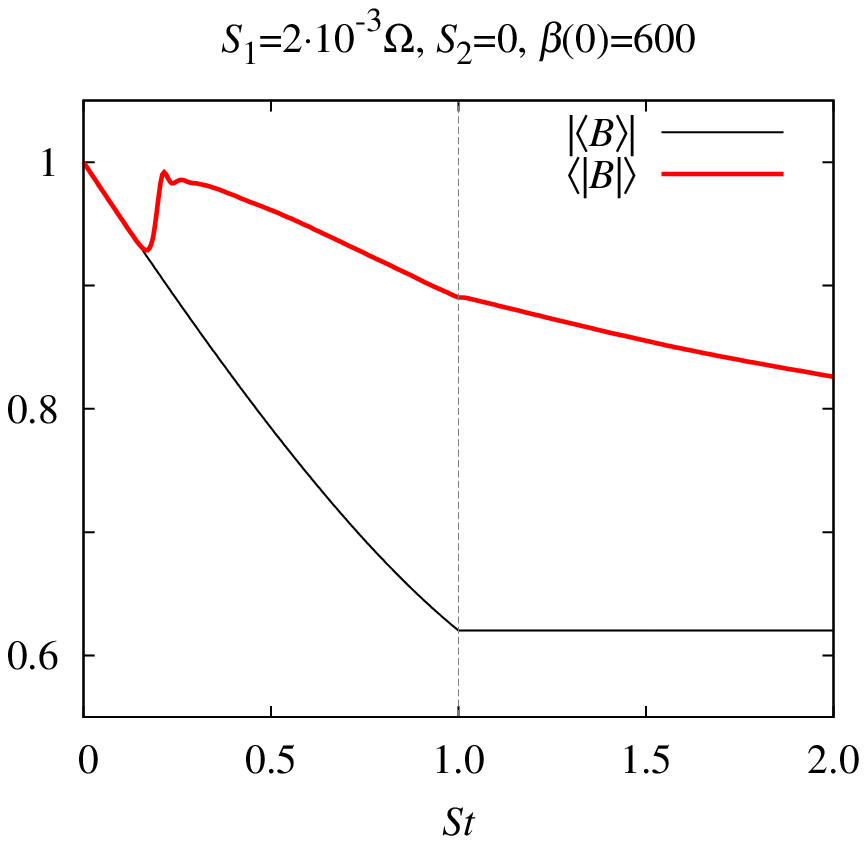}&
\includegraphics[width=0.5\textwidth]{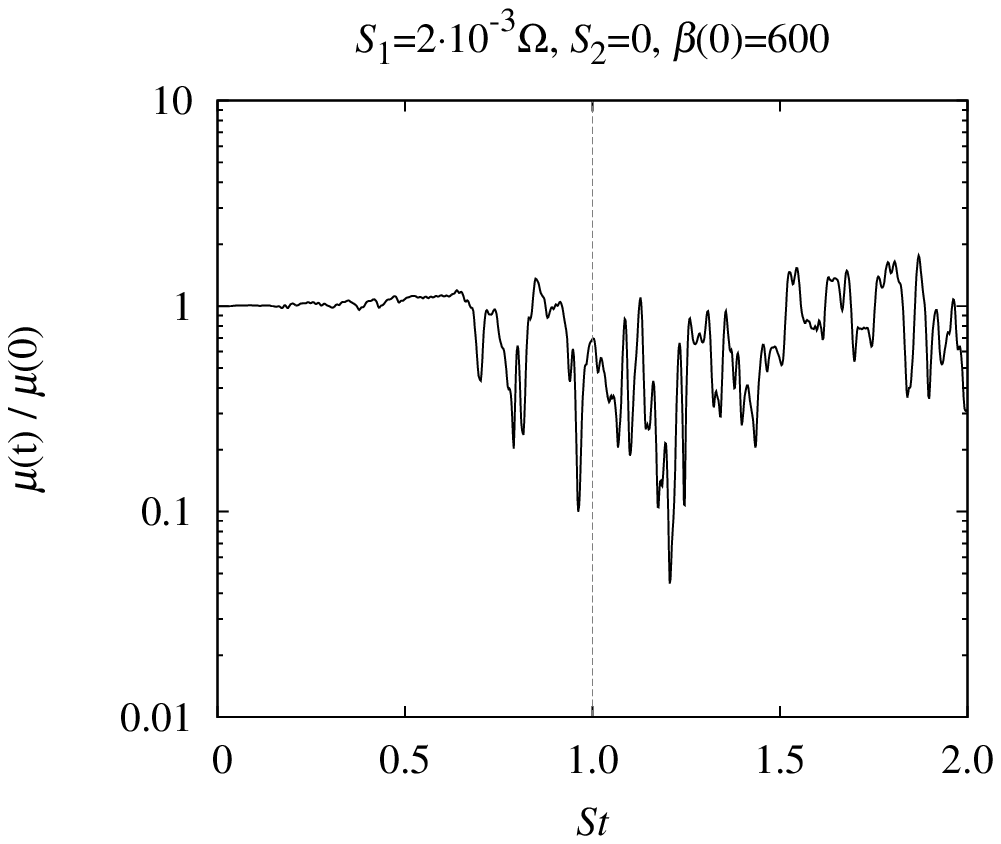}\\
(c)&(d)
\end{tabular}
\caption{Time history of the numerical experiment described in \secref{sec:fh_decay}, 
starting with a shear that decreases the mean field and then switching it off 
after one shear time, $St=1$. In the particular case shown here, 
the initial shear was $S=2\times10^{-3}\Omega$ and $\beta(0)=600$, so this 
case is marginal with respect to the change of physical regime at $\beta\sim\Omega/S$.  
(a) Black solid line: time evolution of $\bsq$. 
The dotted line at $St<1$: the analytical estimate for secular growth (second line of \eqref{eq:fh_sec}, 
where $\vB_0(0)$ is in fact taken at $St=0.25$, the beginning of the secular-growth stage). 
The dotted line at $St>1$: the slope corresponding to the exponential decay at the rate 
$\gamma=0.9\Omega/\beta$ (calculated at $St=1$), consistent with \eqref{eq:decay}. 
Red crosses: the effective collisionality 
$\nueff/\Omega$, calculated by the method described in \apref{ap:coll}
(numerical values are shown on the vertical axis on the right of the plot).
During secular growth, $\nueff$ starts increasing when the fluctuation level approaches $\bsq\sim1$.  
During the decay, $\nueff$ manifestly follows $\bsq$, in agreement with \eqref{eq:nueff_decay}. 
(b) Time evolution of the pressure anisotropy $\Delta=(\pperp-\ppar)/p$. 
Red dotted line shows the marginal level, $\Delta=-2/\beta(t)$ (this 
includes the slow increase of $\beta$ with time during the sheared stage).
The firehose turbulence is near-marginal in all nonlinear growth and decay stages. 
(c) Time evolution of the mean-field magnitude, $B_0=|\la\vB\ra|$ (black thin line) 
and of the mean magnitude of the field, $\la B\ra = \la|\vB|\ra$ (red bold line).
(d) Time evolution of the first adiabatic invariant $\mu=\vperp^2/B$ for a single 
particle: $\mu$ is initially conserved during secular growth, but this 
conservation is broken as fluctuations become large, $\bsq\sim1$, and continues to 
be broken during the decay of the firehose turbulence.}
\label{fig:fh_decay}
\end{figure*}

\subsubsection{Numerical results}
\label{sec:fh_growth_num}

The numerical study by \citet{Kunz14} has shown that both of these scenarios do occur, 
sequentially: in their simulations, after a period of secular growth obeying 
\eqref{eq:fh_sec} and no particle scattering, the (oblique) firehose turbulence 
developed an effective collisionality very close to that given by \eqref{eq:fh_nueff}
and saturated at a level that scaled~as\footnote{There is 
currently no theoretical understanding of this scaling.} 
$\bsq\propto S^{1/2}$. However, these simulations were all done at one 
value of (initial) beta $\beta(0)=200$. In what follows, the $\beta$ dependence of the saturated 
state will matter and so we now extend the \citet{Kunz14} parameter study to a range 
$\beta(0)\in[100,1000]$. 

We use the same set up and the same code 
as \citet{Kunz14} and carry out a series of hybrid (kinetic ions, 
isothermal electrons), $\delta\! f$ PIC simulations of a collisionless 
plasma in a 2D box where an initial magnetic field in the $(x,y)$ plane
(with $B_{0x}>0$, $B_{0y}(0)>0$) is slowly reduced by linear shearing as per \eqref{eq:gamma0} 
with $S>0$. The numerical details are documented in \apref{ap:code}. 

We find that firehose fluctuations grow secularly according to 
\eqref{eq:fh_sec} (\figref{fig:fh_growth}) 
until they saturate at a level that scales with both $S$ and $\beta$, 
a new result: 
\beq
\bsq \sim \lt(\beta\,\frac{S}{\Omega}\rt)^{1/2} 
\label{eq:fh_sat}
\eeq
(see \figref{fig:fh_sat}a). 
This scaling can of course only be valid provided the resulting fluctuation level 
is small ($<1$). At sufficiently high $\beta$ and/or $S$, {\em viz.}, such that    
\beq
\beta \gtrsim \frac{\Omega}{S}, 
\eeq
this is no longer true and $\bsq$ saturates at an order-unity level independent 
of either $\beta$ or $S$.
Whereas the secularly growing firehose fluctuations do not 
scatter particles, the saturated ones do, at a rate that scales with 
$S\beta$ in a manner consistent with \eqref{eq:fh_nueff} 
(\figref{fig:fh_sat}b; how the effective 
collisionality is calculated is explained in \apref{ap:coll}). 

Clearly, when $\beta\ll\Omega/S$, reaching saturation requires 
only a fraction of the shear time: 
\beq
\bsq\sim S\tsat \sim\lt(\frac{\beta S}{\Omega}\rt)^{1/2}
\ \Rightarrow\quad 
\tsat\sim \lt(\frac{\beta}{S\Omega}\rt)^{1/2}\ll S^{-1}.
\eeq 
In this limit, 
since typically local shears in a macroscopic plasma flow will 
change in time at the rates comparable to those shears themselves, 
one can safely consider 
the anomalous collisionality associated with the firehose fluctuations 
to turn on instantaneously, in line with the macroscopic modeling assumption 
that we discussed in \secref{sec:intro}. 
This, however, is no longer the case when $\beta\gtrsim\Omega/S$, as it takes 
$\tsat\sim S^{-1}$ to reach a finite $\bsq$ and for the fluctuations to start 
scattering particles. 
Thus, at these ``ultra-high'' betas, the firehose fluctuations grow secularly 
during much of the shear time, approximately conserve $\mu$ and do not 
scatter particles efficiently.\footnote{Note that 
when they saturate and do start scattering particles (provided they ever reach saturation), 
the effective collision rate \exref{eq:fh_nueff} needed to maintain 
the plasma at the firehose threshold is larger than the cyclotron frequency, and so 
the plasma becomes effectively unmagnetised.} 
The implications of this will be discussed in \secsand{sec:fh_to_mr}{sec:dynamo}. 

\Figsref{fig:fh_pics}a and \ref{fig:fh_pics}b show snapshots of the 
firehose fluctuation field for $\beta(0)=600$ and $S/\Omega=2\times10^{-3}$ 
during the period of secular growth ($St=0.4$, \figref{fig:fh_pics}a), 
when the fluctuations do not scatter particles efficiently,  
and later in the evolution ($St=1.0$, \figref{fig:fh_pics}b), 
when the scattering becomes significant (see \figref{fig:fh_decay}a, which 
documents the evolution of both $\bsq$ and $\nueff$ for the same run). 
While precisely what happens in this later stage is not well understood, it may be a useful 
observation that the firehose structures appear to start shredding---perhaps 
a sign of nonlinear interactions of hydrodynamic (advective), rather 
than quasilinear type (the latter is what controls the secular regime; see \citealt{Rosin11}). 
Small, Larmor and sub-Larmor, scales that are produced this way should be more efficient 
at scattering particles.  

\subsection{Free decay of firehose turbulence}
\label{sec:fh_decay}

Let us now consider what happens if we start in a state that is firehose-marginal, 
with firehose fluctuations that have been secularly growing 
for one shear time, $St=1$ (and, depending on $\beta$, might have saturated),  
and remove the drive, i.e., switch off 
the shear (in \secref{sec:shear}, we will ``generalise'' this to reversing
the shear rather than switching it off). 
How fast will the fluctuations decay? 

\subsubsection{Theoretical expectations}

To answer this question, one must realise that decaying fluctuations imply decreasing 
mean magnetic energy, which by itself can drive negative mean pressure anisotropy, 
pushing the system towards an unstable state. It is then reasonable to argue that 
the rate of decay of the firehose turbulence 
should not exceed the marginal level set by \eqref{eq:fh_marg}, but with 
the mean field no longer changing, $\rmd\ln B_0/\rmd t=-\tS=0$:
\beq
\frac{1}{2}\frac{\rmd\bsq}{\rmd t} = -\frac{2\nueff}{\beta}. 
\label{eq:fh_noshear_marg}
\eeq  
Physically, the negative pressure anisotropy generated by the decay 
of the initial fluctuations should in turn generate new 
firehose fluctuations, slowing down the decay to the marginal 
level.\footnote{A disclaimer is in order. \Eqsref{eq:fh_marg}, 
\exref{eq:fh_noshear_marg} and similar ones in what follows 
should not be taken as precise statements: technically speaking, they can 
only be derived as controlled approximations if the scale of the firehose fluctuations 
is large compared to the Larmor scale \citep{Rosin11}. In reality, 
their scale is of the same order as the ion Larmor scale \citep{Kunz14}, 
which is why they are able to scatter particles. As we will see 
in analysing our numerical experiments below, the $\beta$ scalings that we obtain
in this non-rigorous way are nevertheless correct.}

The anomalous collisionality $\nueff$ is itself obviously a function 
of the magnitude of the fluctuations. 
The simplest (even perhaps simplistic) way to estimate this dependence is as follows.  
Consider a gyrating particle streaming along a perturbed field line 
whose parallel scale length is $\lpar$. 
In the limit $\lpar\gg\rho_i$, i.e., when the perturbations 
decorrelate on parallel scales much longer than the Larmor radius 
$\rho_i=\vth/\Omega$ and, therefore, as the particle 
travels the distance $\sim\vth/\Omega$ during one Larmor period, 
the magnetic field it sees remains unchanged, the first adiabatic 
invariant $\mu=\vperp^2/B$ is conserved with exponential precision 
for each particle \citep{Kruskal58}. 
In contrast, if $\lpar\sim\rho_i$ (as it indeed 
is for the fastest-growing oblique firehose perturbations; see \citealt{Kunz14}), 
particle orbits fail to close precisely on themselves 
and a particle's perpendicular momentum, or, equivalently, its $\mu$, 
suffers a random change over each Larmor period $\delta t\sim\Omega^{-1}$: 
\beq
\frac{\Delta\mu}{\mu}\sim\frac{\Delta\vperp}{\vperp}
\sim \frac{\delta F\delta t}{m\vperp} \sim \frac{\dBperp}{B_0},
\label{eq:dmu}
\eeq 
where $\delta F \sim e\vperp\dBperp/c = m\vperp\Omega\,\dBperp/B_0$ is the perturbed 
Lorentz force on the particle.\footnote{There is also a contribution to 
$\delta\mu/\mu$ due to the variation in the field strength, but for firehose 
perturbations, this is second-order, $\dB/B_0\sim\dBperp^2/B_0^2$ (see \eqref{eq:dB_fh}), 
and so can be ignored here.} Assuming diffusive accumulation of these changes---essentially, 
pitch-angle diffusion---the diffusion coefficient is the effective collisionality 
of the plasma permeated by firehose fluctuations:
\beq
\nueff\sim \frac{\la(\Delta\mu/\mu)^2\ra}{\Delta t} \sim \Omega\bsq, 
\label{eq:nueff_decay}
\eeq
where $\Delta t \sim \lpar/\vth\sim\rho_i/\vth=\Omega^{-1}$ is the 
time for the particle to transit through one correlation 
length of the firehose turbulence and so the time 
between random kicks of the size given by \eqref{eq:dmu}. 

Using \eqref{eq:nueff_decay} and 
ignoring the slow time dependence of $\beta$ and $\Omega$ 
due to changing total $B$, we infer from \eqref{eq:fh_noshear_marg} 
an exponential decay of the firehose fluctuations: 
\beq
\bsq \propto e^{-\gamma t},\quad
\gamma \sim \frac{\Omega}{\beta}.
\label{eq:decay} 
\eeq

\begin{figure*}
\begin{tabular}{cc}
\includegraphics[width=0.45\textwidth]{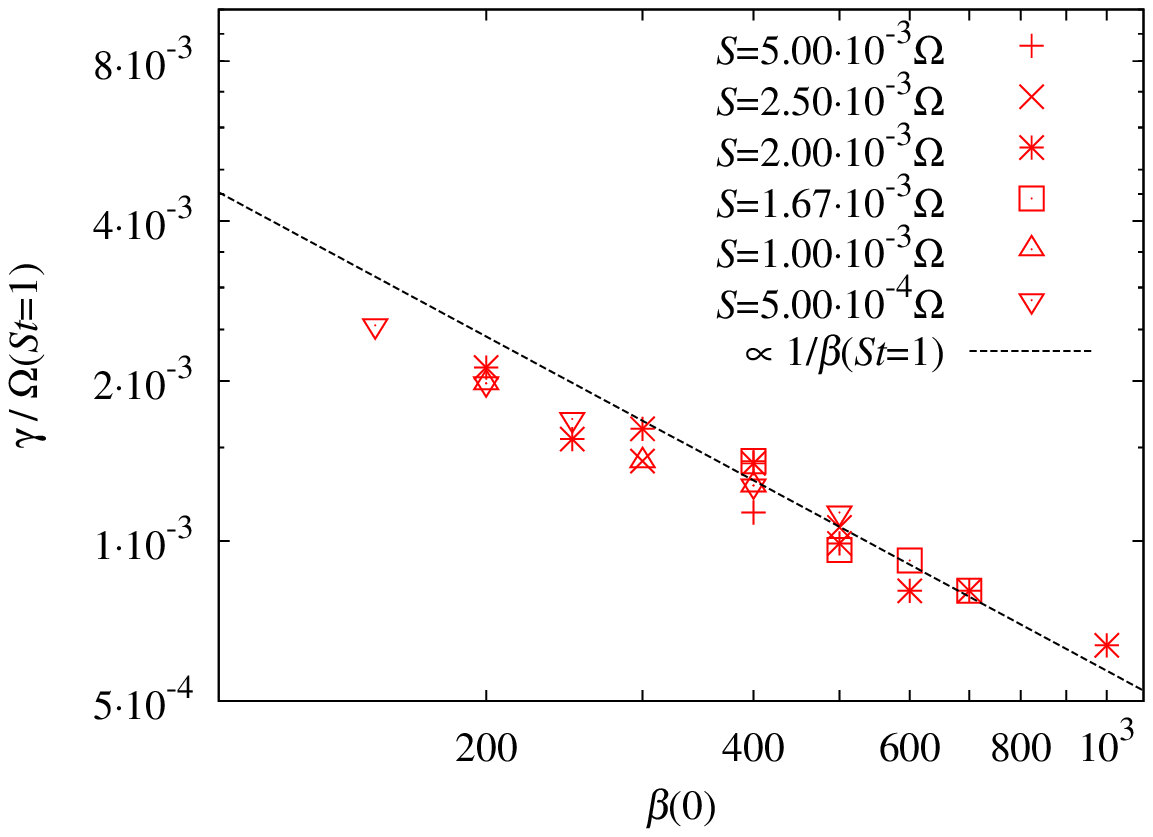}
&\includegraphics[width=0.45\textwidth]{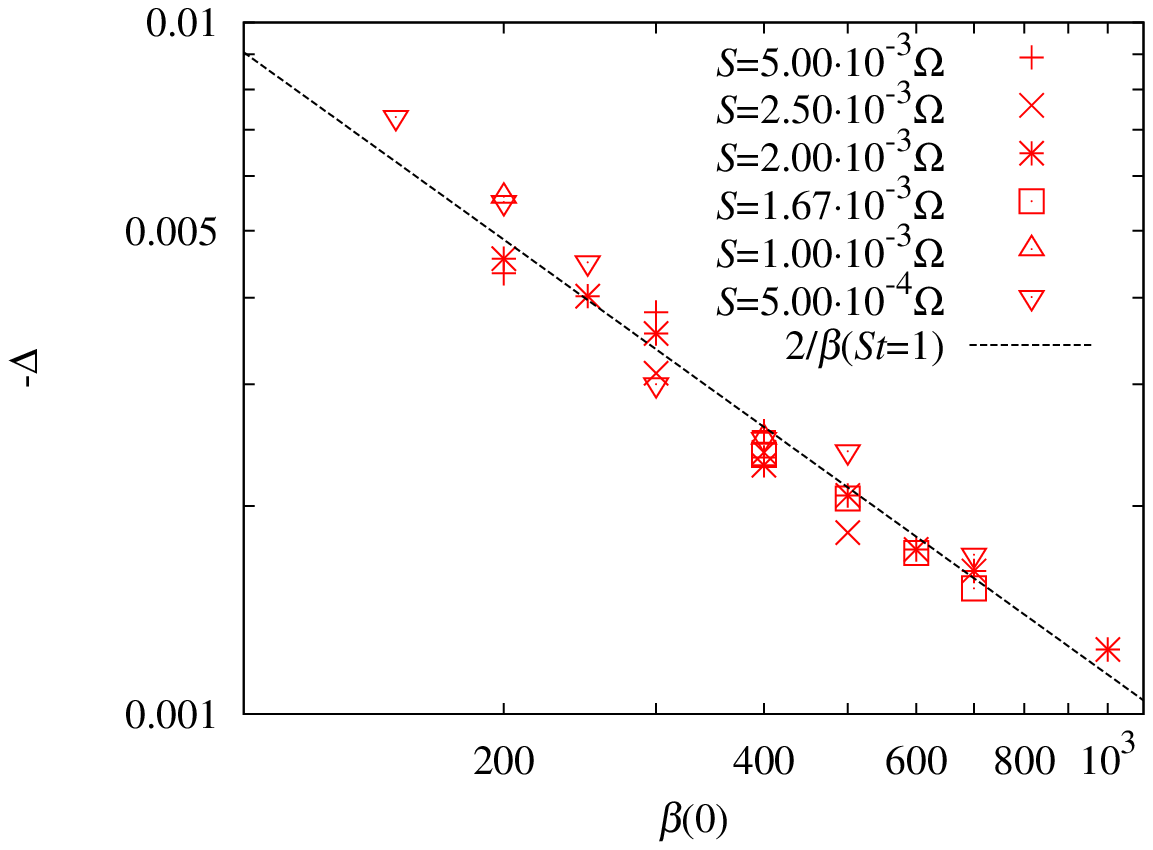}\\
(a)&(b)
\end{tabular}
\caption{(a) Decay exponent $\gamma/\Omega$ for the firehose turbulence 
(\secref{sec:fh_decay}) vs.\ $\beta(0)$, at various 
values of initial $S$ (on which it manifestly does not depend). The fit 
$\gamma = \Omega/\beta(St=1)$ 
is shown as a dotted line and is manifestly 
a good description of the data, confirming \eqref{eq:decay}.
(b) Mean (negative) pressure anisotropy $-\Delta=(\ppar-\pperp)/p$ 
for the same decaying runs. The firehose-marginal level, $-\Delta=2/\beta(St=1)$, 
is shown as a dotted line, confirming that decaying firehose turbulence 
stays marginal, as argued in \secref{sec:fh_decay}.}
\label{fig:gammaDelta_fh_decay}
\end{figure*}

\subsubsection{Numerical results}
\label{sec:fh_decay_num}

Remarkably, this simple argument appears to describe quite well what 
happens in the following numerical experiment. 
For a number of values of $S$ and initial beta $\beta(0)$,  
we first run the same simulations as described in \secref{sec:fh_sat} 
for one shear time; as we explained there, for $\beta\ll\Omega/S$,
this is long enough to obtain saturated, particle-scattering firehose 
turbulence in the marginal state \exref{eq:Delta_fh_marg}, whereas  
for $\beta\gtrsim\Omega/S$, the firehose turbulence 
with $\bsq\sim 1$ is just about formed after one shear time 
(the subsequent evolution that we are about to describe is the same in both regimes). 
Then we switch off the shear and allow the fluctuations to decay. 
This time history is shown in \figref{fig:fh_decay} 
for a representative run (see \figref{fig:fh_pics}a--c for 
snapshots of the firehose fluctuation field for this run 
at stages 2, 3 and 4 described below). 
The system goes through the following stages of evolution: 

\begin{enumerate}

\item[1.]
initial, linear growth of firehose fluctuations as the pressure 
anisotropy, starting from zero, crosses the threshold \exref{eq:fh}, 
overshoots\dots 

\item[2.]
\dots and is then returned to the marginal level \exref{eq:Delta_fh_marg}
as linear growth of fluctuations gives way to secular growth, according to
\eqref{eq:fh_sec}; there is no particle scattering until\dots

\item[3.]
\dots the fluctuation energy approaches 
the level given by \eqref{eq:fh_sat} or by $\bsq\sim1$ if $\beta\gtrsim\Omega/S$, 
giving rise to $\nueff\sim S\beta$, as per \eqref{eq:fh_nueff};\footnote{In the 
ultra-high-$\beta$ regime, while the onset of anomalous scattering occurs 
as the fluctuation level approaches unity, $\bsq\sim1$, this does 
not stop $\bsq$ from continuing to increase (although this increase is partly 
due to $B_0(t)$ decreasing slowly, as we have defined $\bsq = \la|\dvBperp|^2\ra/B_0^2(t)$ 
with respect to the time-dependent mean field).} 

\item[4.]
the shear is switched off at $St=1$ and fluctuations decay exponentially, 
at a rate given by \eqref{eq:decay}; during this decay, 
the effective collisionality $\nueff$
scales with $\bsq$ according to \eqref{eq:nueff_decay}, while the pressure 
anisotropy stays firehose-marginal.\footnote{Note that when the shear is switched 
off, the anisotropy first relaxes very quickly to zero (see \figref{fig:fh_decay}b). 
The time scale on which this happens is consistent with the collision time associated 
with the firehose fluctuations at that moment: from \figref{fig:fh_decay}a, 
$S/\nueff\sim0.01$. The subsequent return to a firehose-marginal state occurs 
from the stable side (i.e., $\Delta\to-2/\beta$ from above). This means that 
the decaying firehose fluctuations, unlike the driven ones, have some similarity 
to propagating Alfv\'en waves, albeit with reduced phase speed \citep[cf.][]{Kunz15}. 
\label{fn:fh_decay}} 

\end{enumerate} 

\begin{figure}
\includegraphics[width=0.45\textwidth]{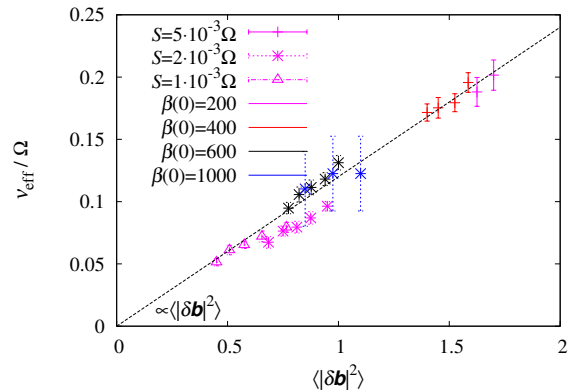}
\caption{Effective collisionality $\nueff/\Omega$ vs.\ mean square fluctuation level $\bsq$
at various times during the decay stage of the firehose turbulence (\secref{sec:fh_decay}) 
and at various values of initial $S$ and $\beta$. 
The dotted line shows the fit $\nueff/\Omega=0.15\bsq$, consistent with the scaling 
\exref{eq:nueff_decay} and manifestly a good description of the data.} 
\label{fig:nueff_fh_decay}
\end{figure}

\Figref{fig:gammaDelta_fh_decay}a confirms that the decay rate scales with $\beta$ 
according to \eqref{eq:decay} and is independent of the initial shear 
$S$ for a number of values of these parameters. \Figref{fig:gammaDelta_fh_decay}b 
confirms that the pressure anisotropy (which is approximately constant in the 
decay stage of all these runs) always stays at the 
firehose-marginal level \exref{eq:Delta_fh_marg}. Finally, 
\figref{fig:nueff_fh_decay} shows that the effective collisionality 
scales with $\bsq$ as predicted by \eqref{eq:nueff_decay} at all values 
of $S$ and $\beta$ and also at various times during the decay. 

\subsection{Growth and saturation of mirror fluctuations}
\label{sec:mr_growth}

Let us now consider what happens when the mean field is locally 
increased, rather than decreased, driving the system through the mirror threshold \exref{eq:mirr}. 
Compared to the firehose, this is an analytically more complicated situation, 
a careful treatment of which can be found in \citet{Rincon15}. 
Here, we will give a vulgarised qualitative discussion, at the risk of 
outraging a purist reader, but hopefully with the benefit of enlightening 
an impatient one. 

\begin{figure*}
\begin{tabular}{cc}
\includegraphics[width=0.45\textwidth]{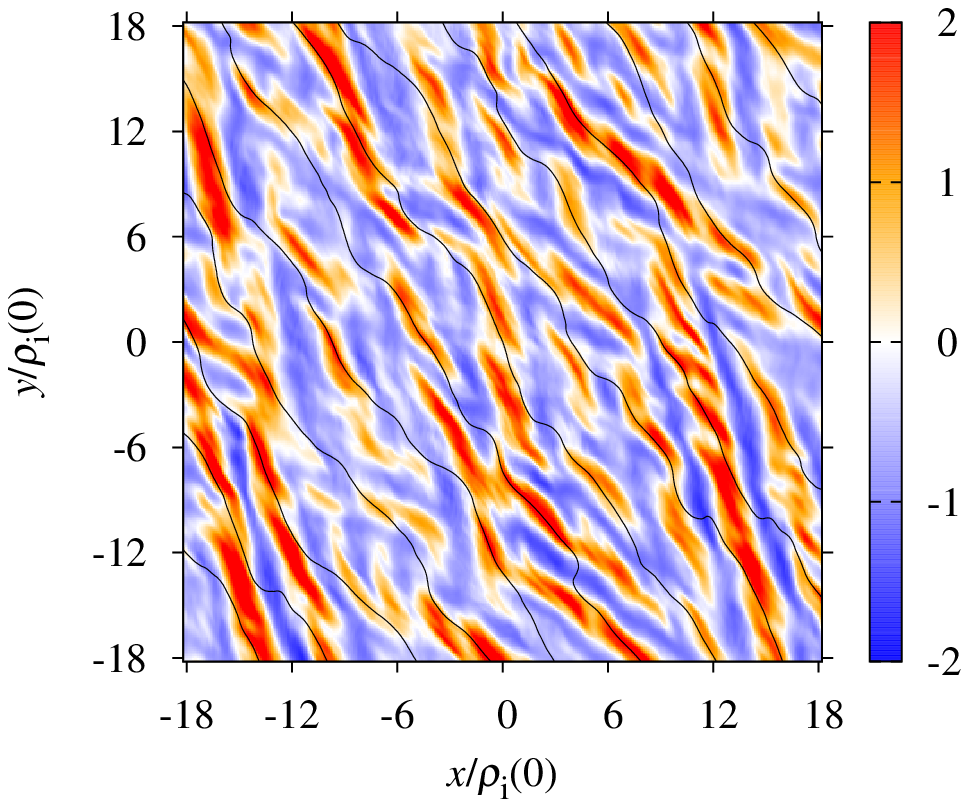}&
\includegraphics[width=0.45\textwidth]{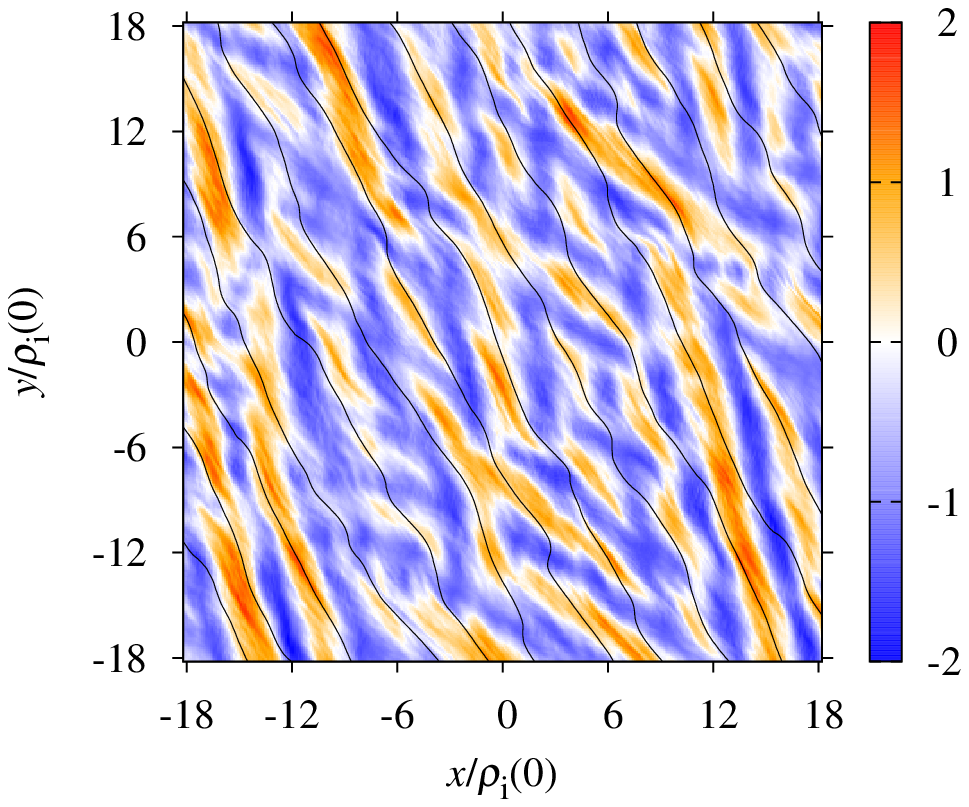}\\
(a) Growing mirror, $\dBpar(x,y)$ at $|S|t = 1.0$ & 
(b) Decaying mirror, $\dBpar(x,y)$ at $|S|t = 1.4$\\\\
\includegraphics[width=0.45\textwidth]{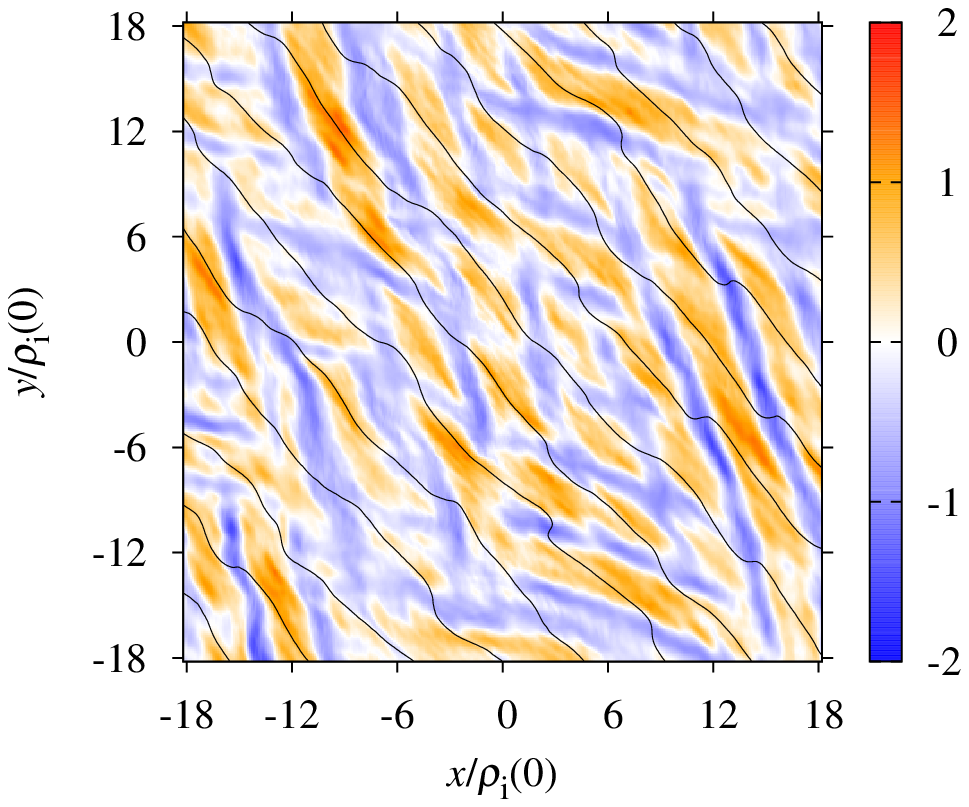}&
\includegraphics[width=0.45\textwidth]{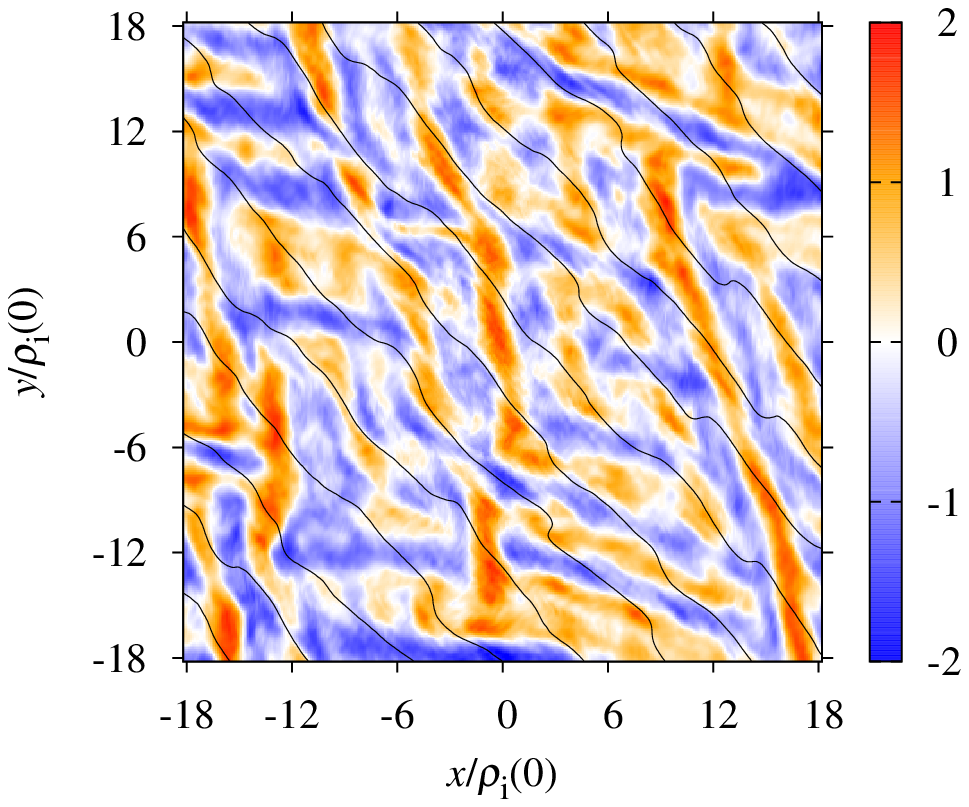}\\
(c) From mirror to firehose, $\dBpar(x,y)$ at $|S|t = 1.4$ & 
(d) From mirror to firehose, $\dB_z(x,y)$ at $|S|t = 1.4$
\end{tabular}
\caption{
A series of snapshots of the fluctuating fields in simulations 
with $\beta(0)=600$ and $S=-2\times10^{-3}$,  
starting in the mirror-unstable regime: (a) $\dBpar(x,y)$ 
at $|S|t=1.0$ (growing mirror perturbations; see \figref{fig:mr_decay}a and d); 
(b) same as (a) at $|S|t=1.4$ in a run where shear was switched off at $|S|t=1$ 
(decaying mirror perturbations; see \figref{fig:mr_decay}a));
(c) same as (a) at $|S|t=1.4$ in a run where shear was reversed at $|S|t=1$
(mirror perturbations being sheared away; see \figref{fig:mr_to_fh}a);
(f) $\dB_z(x,y)$ (proxy for $\dvBperp$) in the same run and at the same time as (c) 
(growing firehose perturbations; see \figref{fig:mr_to_fh}a). 
The $x$ and $y$ coordinates are shown in the units of the ion Larmor
radius at the start of the simulation, $\rho_i(0)$. 
Only the central $36\times36\rho_i^2(0)$ patch of a larger 
$82\times82\rho_i^2(0)$ simulation box is shown, for 
comparison with \figref{fig:fh_pics}. 
Black lines are magnetic-field lines. 
Note that the range of amplitudes is large, $\dBpar\in[-2,2]$, because 
they are normalised to the initial field 
(whereas the mean field $B_0(t)$ grows with time at $|St|\le1$).} 
\label{fig:mr_pics}
\end{figure*}

Let us continue using \eqref{eq:Delta} to describe 
the evolution of the mean pressure anisotropy, except the angle averaging 
will now mean averaging over particle trajectories. For passing particles, 
this is the same as spatial averaging, but, in the case of mirror 
fluctuations, there will also be a population of particles trapped 
in these fluctuations---for these particles, the average 
that matters in the calculation of the pressure anisotropy 
is the bounce average \citep{Rincon15}. 
The mirror fluctuations are, to lowest order in their amplitude, 
fluctuations of the magnetic-field strength, or $\dBpar$ 
(a snapshot showing what they look like is given in \figref{fig:mr_pics}a
for a run with the same parameters as the firehose run in 
\figref{fig:fh_pics}, but with the opposite sign of the shear). 
Therefore,
\beq
\la\ln B\ra \approx \ln B_0 + \lt\la\frac{\dBpar}{B_0}\rt\ra 
\sim \ln B_0 - \lt|\frac{\dBpar}{B_0}\rt|^{3/2},
\eeq
where the last, qualitative, estimate is based on the expectation that 
the only non-zero contribution to $\la\dBpar\ra$ will come from 
the trapped particles, that it is negative because the trapped 
particles are trapped in areas of smaller $B$, and that the 
fraction of these particles in the system is $\sim |\dBpar/B_0|^{1/2}$. 
Assuming that the pressure anisotropy will tend to marginal,
\beq
\Delta\to\frac{1}{\beta},
\label{eq:Delta_mr_marg}
\eeq
the analogue of \eqref{eq:fh_marg} for the mirror is 
(non-rigorously)
\beq
\frac{\rmd\ln B_0}{\rmd t} - \frac{\rmd}{\rmd t}\lt|\frac{\dBpar}{B_0}\rt|^{3/2}
\sim \frac{\nueff}{\beta}.
\label{eq:mr_marg}
\eeq
Again, in view of \eqref{eq:gamma0}, there is a secular growth regime \citep{Sch08}, 
\beq
\lt|\frac{\dBpar}{B_0}\rt| \sim (|S|t)^{2/3},
\label{eq:mr_secular}
\eeq
assuming inefficient scattering of particles by mirror modes ($\nueff\ll S\beta$)
and so an approximate balance in the left-hand side of \eqref{eq:mr_marg}
(for the rigorous derivation of this secular evolution, see \citealt{Rincon15}). 

\begin{figure*}
\begin{tabular}{cc}
\includegraphics[width=0.5\textwidth]{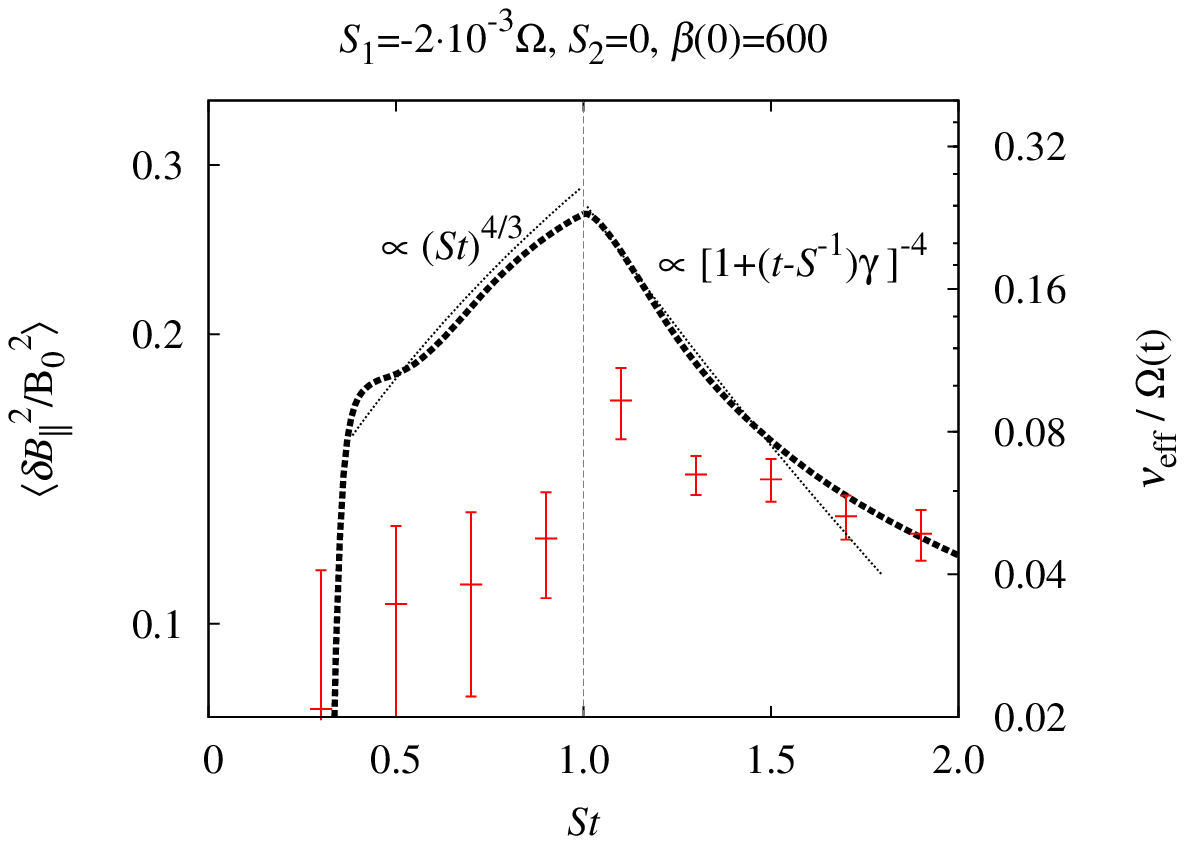}&
\includegraphics[width=0.5\textwidth]{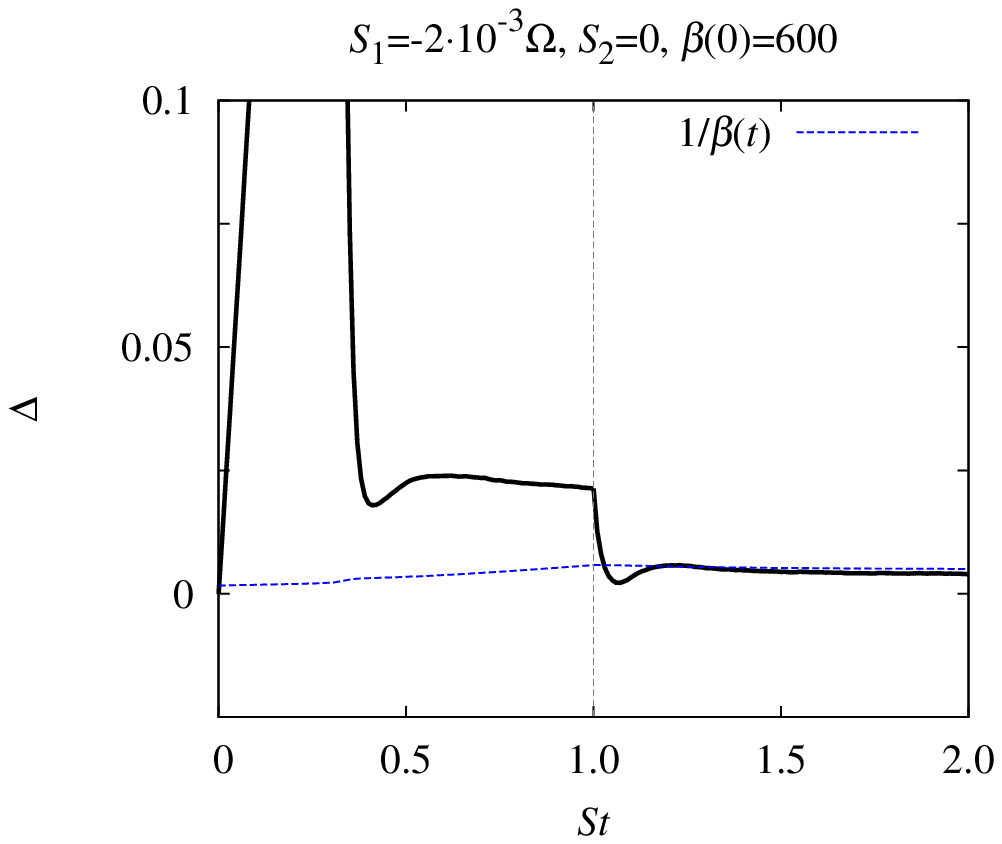}\\
(a)&(b)\\
\includegraphics[width=0.5\textwidth]{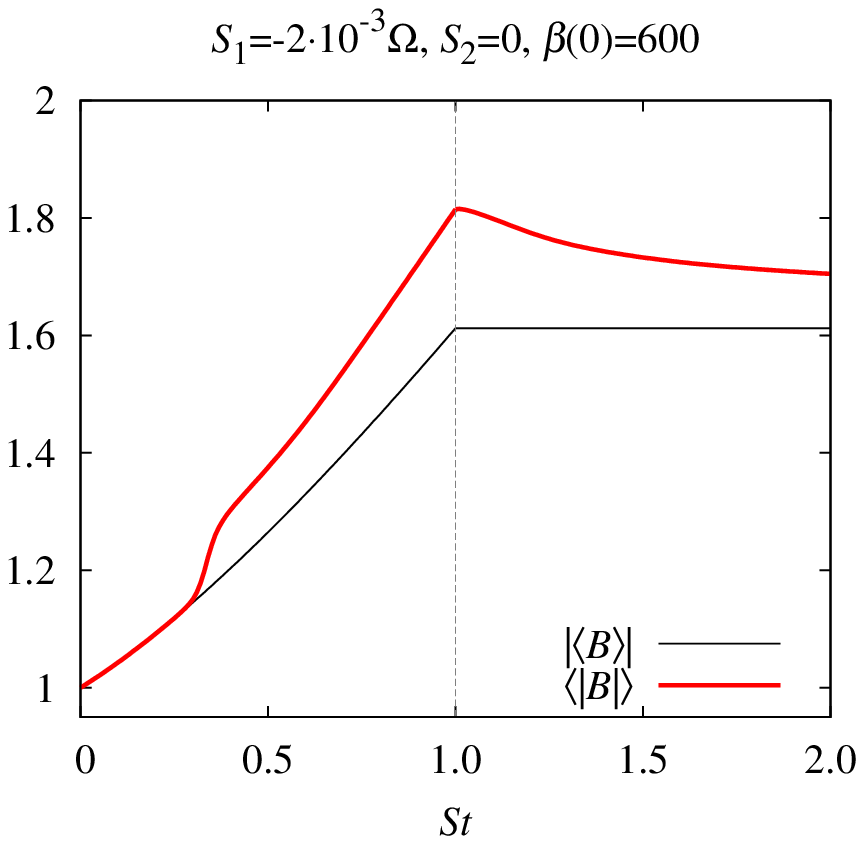}&
\includegraphics[width=0.5\textwidth]{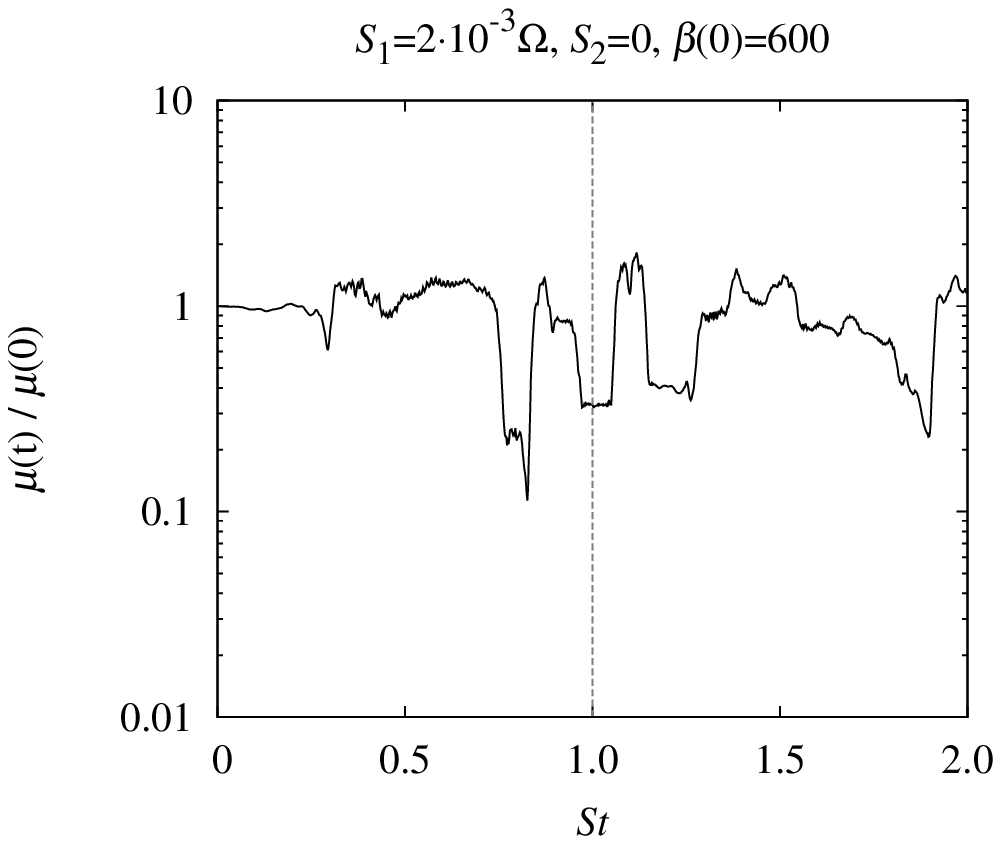}\\
(c)&(d)
\end{tabular}
\caption{Same as \figref{fig:fh_decay}, but starting with negative shear, 
$S=-2\times10^{-3}\Omega$, and so increasing the mean magnetic field. 
Same quantities are shown, with the following differences.
(a) Time evolution of the mirror perturbations, $\la\dBpar^2/B_0^2\ra$
is shown by the bold black dotted line; time-dependent mean field $B_0(t)$ 
is used for normalisation. 
Dotted lines show the theoretical estimates for the secular 
growth of driven mirror turbulence at $|S|t<1$, \eqref{eq:mr_secular}, 
and for the decay of the decaying one at $|S|t>1$, \eqref{eq:mr_decay}
(with $t_0=|S|^{-1}$ and $\gamma = 0.19\Omega/\beta$, calculated at $St=1$; 
note that the label in the plot omits the factor of 
$[\la\dBpar^2/B_0^2\ra(t_0)]^{1/4}$ in from of $\gamma$ 
for lack of space, but this factor \emph{is} present in the fit).
(b) Blue dotted line shows the marginal level of pressure anisotropy for the  
mirror instability, $\Delta=1/\beta(t)$. 
(d) The first adiabatic invariant $\mu=\vperp^2/B$ 
is conserved throughout the sheared stage.} 
\label{fig:mr_decay}
\end{figure*}

This is indeed what happens for 
much of the nonlinear evolution of the mirror instability \citep{Kunz14}. 
Physically, just like the firehose fluctuations in the secular-growth regime 
combined to compensate on average for the decline of the mean field and thus 
allow the particles to conserve their $\mu$, the mirror fluctuations 
accomplish a similar feat but by a different ruse: as the mean 
field grows, an increasing number of particles gets trapped in 
increasingly deep mirrors, protecting those particles from the effect 
the growing field would have had on their $\mu$ \citep[cf.][]{Kivelson96,Pantellini98,Pokhotelov08,Rincon15}. 
Thus, whereas both $|\la\vB\ra|$ and $\la|\vB|\ra$ grow in the volume-averaged 
sense, $\la|\vB|\ra$ does {\em not} grow if averaged over 
particle trajectories. \Figref{fig:mr_decay}c documents 
the time evolution of both the volume-averaged (mean) field $|\la\vB\ra|$ and 
the volume-averaged total field strength $\la|\vB|\ra$ during such a secular stage 
(at $|S|t<1$). Unlike in the case of the decaying mean field 
(\figref{fig:fh_decay}c), where the secular growth of the firehose 
fluctuations slowed down the rate of decay of the total field strength, 
the mirror fluctuations do not decrease the rate of the total field's growth. 
This is because marginal pressure anisotropy is achieved by rearranging 
particles between the passing and trapped populations, rather than by impeding 
the change of~$B$.\footnote{Note that this means that the upper bound on the 
growth rate of $B$ conjectured by \citet[][\S 3]{Mogavero14} is likely 
to have been too pessimistic (see updated discussion in \secref{sec:mag}).}

Scattering is small in this regime (see \figsref{fig:mr_decay}a and d at $St<1$, 
and more asymptotic cases with lower $S$ in \citealt{Kunz14}) 
because both linear and nonlinear mirror modes near the threshold are quite 
well separated from the Larmor scale \citep{Hellinger07,Rincon15} 
and thus conserve $\mu$; 
a significant $\nueff$ is only achieved after the mirror turbulence saturates at 
\beq
\lt|\frac{\dBpar}{B_0}\rt| \sim 1,
\eeq 
independently of $S$ \citep{Kunz14,Riquelme15}, with scattering 
likely occurring at sharp (Larmor-scale) boundaries of the mirror wells. 
The time to saturation is $St\sim1$ and so the secular regime should  
be the relevant one most of the time during each episode of the 
mean-field growth. 

\begin{figure*}
\begin{tabular}{cc}
\includegraphics[width=0.45\textwidth]{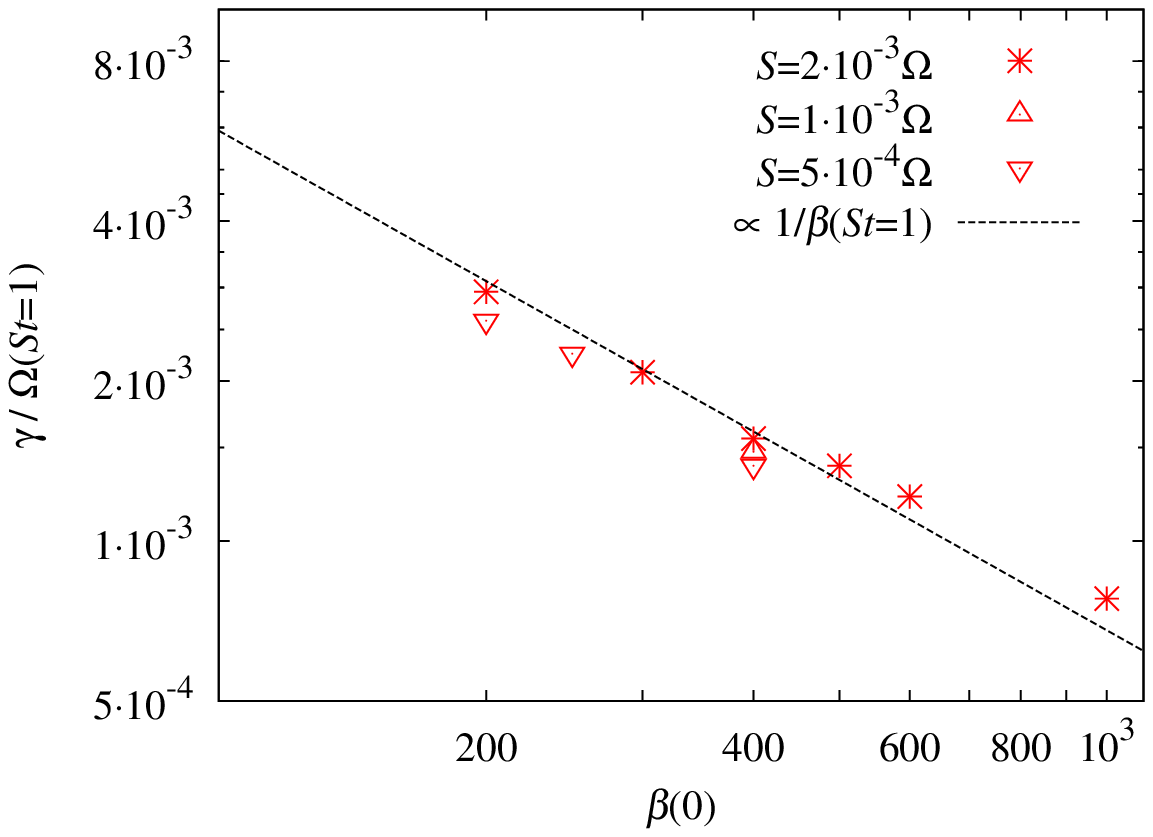}
&\includegraphics[width=0.45\textwidth]{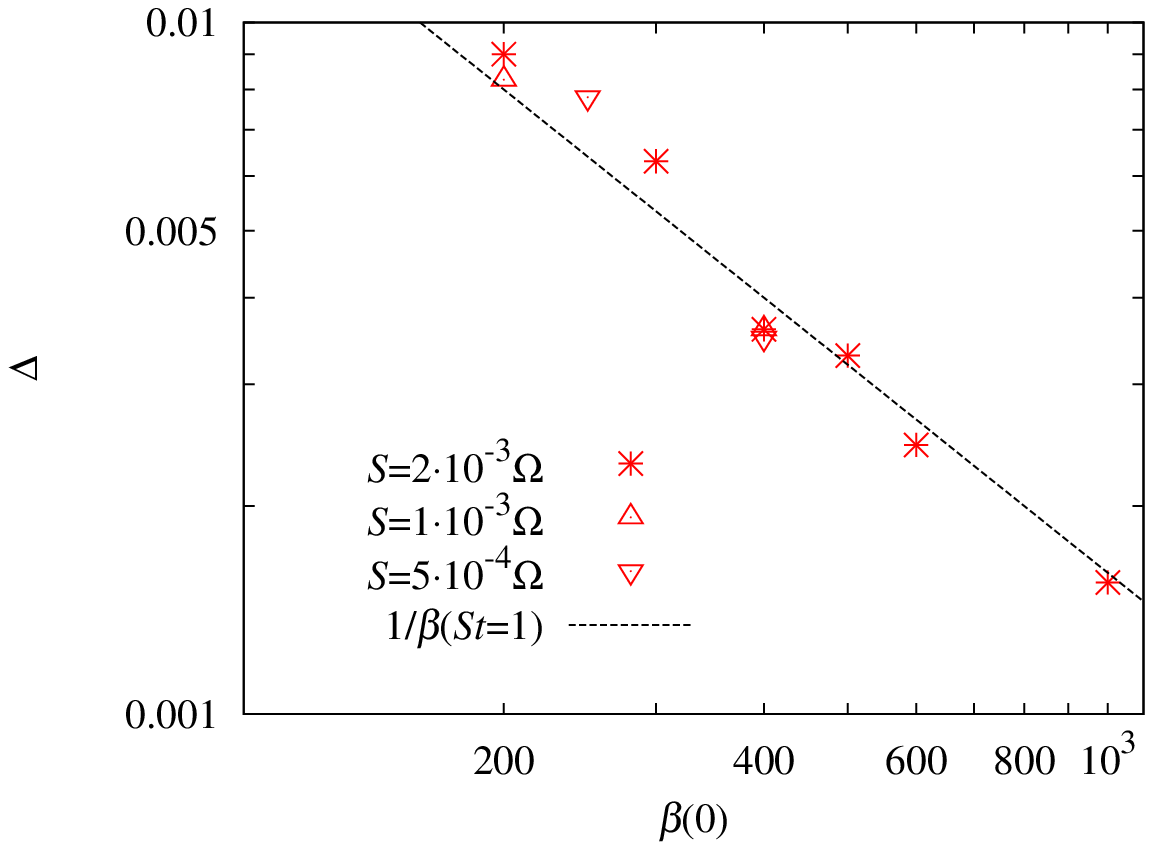}\\
(a)&(b)
\end{tabular}
\caption{(a) Inverse decay time scale $\gamma/\Omega$ for the mirror turbulence, 
based on fitting the data to the decay law \exref{eq:mr_decay},  
vs.\ $\beta(0)$, at various 
values of initial $S$ (on which it manifestly does not depend). The fit 
$\gamma = 0.19\Omega/\beta(|S|t=1)$ is shown as a dotted line and is manifestly 
a good description of the data, supporting \eqref{eq:mr_gamma}.
(b) Mean pressure anisotropy $\Delta=(\ppar-\pperp)/p$ 
for the same decaying runs. The mirror-marginal level, $\Delta=1/\beta(|S|t=1)$, 
is shown as a dotted line, confirming that decaying mirror turbulence 
stays marginal, as argued in \secref{sec:mr_decay}.}
\label{fig:gammaDelta_mr_decay}
\end{figure*}

\subsection{Free decay of mirror turbulence}
\label{sec:mr_decay}

\subsubsection{Theoretical expectations}

Just as we did for the firehose case, we now ask what happens if we 
switch off the shear at $St=1$. The mirrors must start decaying 
and, as they decay, more and more of the particles previously trapped 
in them are released. As these particles escape the mirrors, they are 
able to sample regions of stronger field and so, as far as the particles 
are concerned, they, on average, see an {\em increasing} field and so 
the pressure anisotropy will be pushed upwards! This seemingly 
paradoxical conclusion is made more intuitive by noting that the net effect 
of the secularly growing mirror fluctuations in the driven regime described in 
\secref{sec:mr_growth} was to push the pressure anisotropy down; 
now that these mirror fluctuations decay, the anisotropy will go back up. 
As it is limited from above by the mirror instability threshold, 
it is natural to assume (analogously to what we did in \secref{sec:fh_decay}) 
that the decaying mirror turbulence will stay in the marginal state 
\exref{eq:Delta_mr_marg}. With the drive (growing $B_0$) removed, 
\eqref{eq:mr_marg} becomes
\beq
- \frac{\rmd}{\rmd t}\lt|\frac{\dBpar}{B_0}\rt|^{3/2} 
\sim \frac{\nueff}{\beta}.
\label{eq:mr_noshear_marg}
\eeq

The effective collisionality $\nueff$ can be estimated by assuming 
that the breaking of $\mu$ conservation will happen as particles encounter 
sharp edges of saturated mirrors \citep{Kunz14}. By the argument 
that led to \eqref{eq:dmu}, the relative change in $\mu$ resulting 
from such an encounter is 
\beq
\frac{\Delta\mu}{\mu}\sim\Omega\,\delta t\,\frac{\dBperp}{B_0}
\sim \Omega\,\delta t\,\frac{\lperp}{\lpare}\lt|\frac{\dBpar}{B_0}\rt|,
\label{eq:dmu_mr1}
\eeq 
where $\delta t$ is the time the particle takes to transit the 
sharp-edge region, which must be shorter than the cyclotron period, 
$\Omega\,\delta t \lesssim 1$ 
(otherwise $\mu$ will be conserved with exponential precision; see 
\citealt{Kruskal58}). 
The second expression in \eqref{eq:dmu_mr1} was obtained by using 
solenoidality of the magnetic field to estimate the perpendicular 
perturbation of the magnetic field in a mirror's edge in terms of 
its parallel perturbation; $\lperp$ is the perpendicular scale of 
the mirror and $\lpare$ is the parallel extent of its edge. 
If $\vpar$ is the parallel velocity of the particle, then 
$\delta t \sim \lpare/\vpar$ and we get\footnote{The contribution 
to $\Delta\mu$ due to the change in the denominator of $\mu=\vperp^2/B$, 
$\Delta\mu/\mu\sim |\dBpar/B_0|$, is always smaller than that given 
by \eqref{eq:dmu_mr2} because $\lperp\gg\rho_i$ for near-marginal 
mirrors \citep{Hellinger07,Rincon15}.} 
\beq
\frac{\Delta\mu}{\mu}\sim 
\frac{\lperp}{\rho_i}\lt(\frac{\vpar}{\vth}\rt)^{-1}\lt|\frac{\dBpar}{B_0}\rt|.
\label{eq:dmu_mr2}
\eeq
The prefactor $(\vpar/\vth)^{-1}\sim1$ for the passing particles and 
$(\vpar/\vth)^{-1}\sim |\dBpar/B_0|^{-1/2}$ for the trapped ones; the latter 
suffer a bigger change in $\mu$ because they take longer to transit 
the edge region. Note that the condition $\Omega\,\delta t \lesssim 1$ 
implies that the scale of the sharp region should be  
\beq
\lpare \lesssim \lt|\frac{\dBpar}{B_0}\rt|^{1/2}\rho_i 
\eeq
for the trapped particles to be scattered, enabling 
the mirror modes to saturate at the end of their driven stage. 

Assuming again that the kicks to $\mu$ given by \eqref{eq:dmu_mr2} accumulate 
diffusively, the pitch-angle diffusion coefficient, or the effective collisionality 
of the plasma permeated by mirrors,~is 
\begin{align}
\nonumber
\nueff&\sim\frac{\la(\Delta\mu/\mu)^2\ra}{\Delta t}\sim 
\frac{(\lperp/\rho_i)^2}{\lparm/\rho_i}\,\Omega\lt(\frac{\vpar}{\vth}\rt)^{-1}
\lt|\frac{\dBpar}{B_0}\rt|^2\\
&\sim \Omega\lt(\frac{\vpar}{\vth}\rt)^{-1}
\lt|\frac{\dBpar}{B_0}\rt|^2. 
\label{eq:nueff_mr1}
\end{align}
To obtain this estimate, we have used \eqref{eq:dmu_mr2} and 
assumed that the time between random kicks of the size given by it 
is the time for the particles to transit the full length $\lparm$ 
of a mirror, $\Delta t\sim \lparm/\vpar$ (for the trapped particles, 
this is the bounce time). The final expression in 
\eqref{eq:nueff_mr1} follows if we also assume that we can use for the 
nonlinear mirrors \citep{Kunz14,Rincon15} the result that 
$\lparm/\rho_i\sim (\lperp/\rho_i)^2\sim(\beta\Delta - 1)^{-1}$ 
for the fastest growing linear mirror modes \citep{Hellinger07}.  
Finally, letting $\vpar/\vth\sim1$ for the passing particles 
and $\vpar/\vth\sim|\dBpar/B_0|^{1/2}$ for the trapped ones, we 
arrive~at
\begin{align}
\label{eq:nueff_mr_decay_trapped}
\nueff^\mathrm{(trapped)} &\sim \Omega \lt|\frac{\dBpar}{B_0}\rt|^{3/2},\\ 
\nueff^\mathrm{(passing)} &\sim \Omega \lt|\frac{\dBpar}{B_0}\rt|^2. 
\label{eq:nueff_mr_decay_passing}
\end{align}
The trapped particles are scattered more vigorously. 

Since \eqref{eq:mr_noshear_marg} was deduced (non-rigorously) by accounting  
for the bounce-averaged contribution of the trapped particles to the 
pressure anisotropy (see \eqref{eq:Delta} and \secref{sec:mr_growth}), 
we should use the effective collisionality of the trapped population. 

The effective collisionality that appears in \eqref{eq:mr_noshear_marg}
is the rate of relaxation of pressure anisotropy, to which the scattering 
of both trapped and passing particles contributes. Therefore, the total 
collisionality must be an average of $\nueff^\mathrm{(trapped)}$ 
and $\nueff^\mathrm{(passing)}$ weighted by the fraction of each type 
of particles. Since the fraction of trapped particles is $\sim|\dBpar/B_0|^{1/2}$, 
this weighted average is 
\beq
\label{eq:nueff_mr_decay}
\nueff \sim \Omega \lt|\frac{\dBpar}{B_0}\rt|^2.
\eeq
Substituting this formula into \eqref{eq:mr_noshear_marg}, we get
\beq
\frac{\rmd}{\rmd t}\lt|\frac{\dBpar}{B_0}\rt|^{3/2} 
\sim -\gamma \lt|\frac{\dBpar}{B_0}\rt|^2,
\quad \gamma \sim\frac{\Omega}{\beta}. 
\label{eq:mr_gamma}
\eeq
The solution for the mean square mirror fluctuation is 
\beq
\lt\la\frac{\dBpar^2}{B_0^2}\rt\ra\!(t) = 
\frac{\la\dBpar^2/B_0^2\ra(t_0)}{\lt\{1 + (t-t_0)\gamma\lt[\la\dBpar^2/B_0^2\ra(t_0)\rt]^{1/4}\rt\}^4},
\label{eq:mr_decay}
\eeq
where $t_0$ is the time at which free decay starts, in our case $t_0=|S|^{-1}$. 
Thus, mirrors decay on the same time scale $\sim\gamma^{-1}$ 
(assuming $\dBpar/B_0\sim1$ at $t=t_0$) 
as firehose fluctuations did in \secref{sec:fh_decay}, but the decay is slower than exponential.  

\begin{figure}
\includegraphics[width=0.45\textwidth]{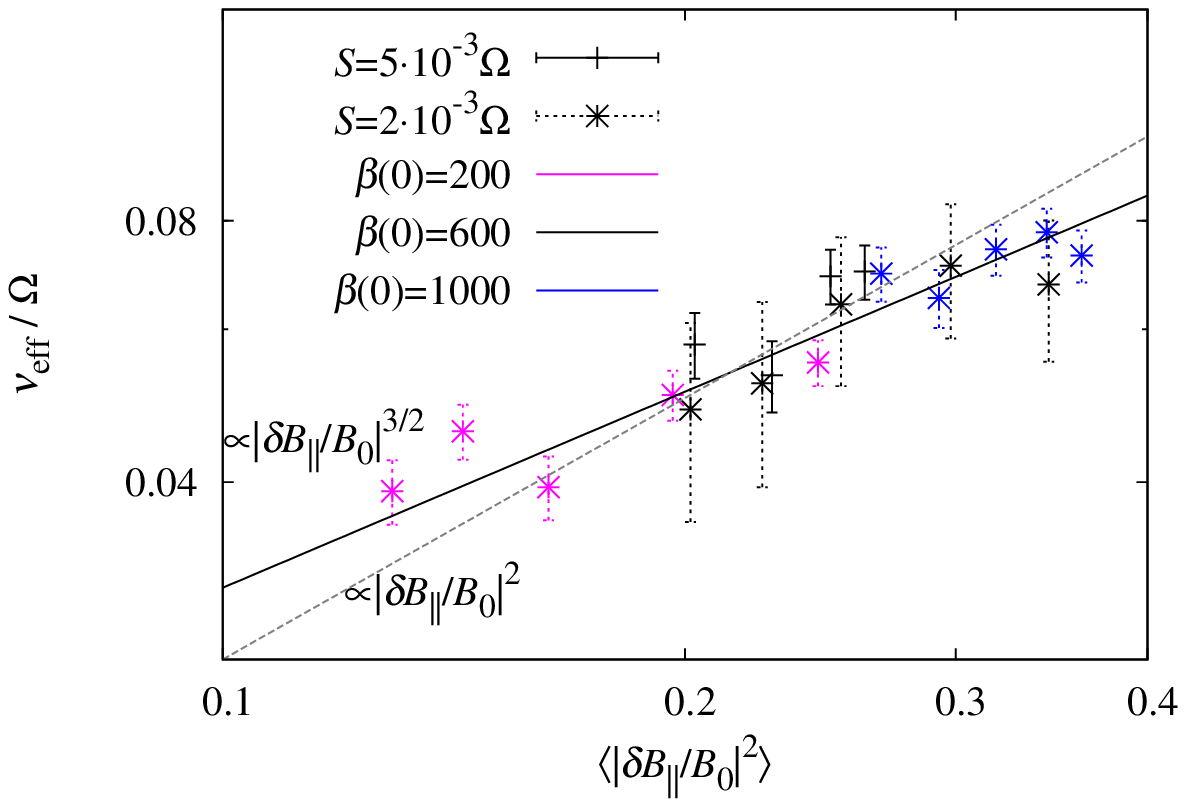}
\caption{Effective collisionality $\nueff/\Omega$ vs.\ $\la|\dBpar/B_0|^2\ra$
(mean over the box) 
at various times during the decay stage for the mirror turbulence (\secref{sec:mr_decay}) 
and at various values of initial $S$ and $\beta$. 
The solid straight line shows the fit $\nueff/\Omega = 0.17\la|\dBpar/B_0|^2\ra^{3/4}$
and the dashed one $\nueff/\Omega = 0.25\la|\dBpar/B_0|^2\ra$, 
corresponding to the scalings \exref{eq:nueff_mr_decay_trapped} 
and \exref{eq:nueff_mr_decay_passing}, respectively.} 
\label{fig:nueff_mr_decay}
\end{figure}

\subsubsection{Numerical results}
\label{sec:mr_decay_num}

\Figref{fig:mr_decay} shows the time history of a numerical experiment 
analogous to the one described in \secref{sec:fh_decay}, except starting 
in a mirror-marginal state populated by mirror fluctuations. 
A snapshot of the mirror fluctuation field in free decay is shown in 
\figref{fig:mr_pics}b. 

In line with the above arguments, we see mirror fluctuations 
decay consistently with \eqref{eq:mr_decay}, at least for a reasonable 
initial period (\figref{fig:mr_decay}a)---although at the level of precision 
and asymptoticity that we can afford in these simulations, we cannot 
distinguish definitively between the decay law \exref{eq:mr_decay} and 
an exponential decay $\la\dBpar^2/B_0^2\ra\propto e^{-\gamma t}$. 
The decay time $\gamma^{-1}$ scales as predicted in \eqref{eq:mr_gamma} 
(see \figref{fig:gammaDelta_mr_decay}a), 
while the pressure anisotropy stays mirror-marginal (\figref{fig:mr_decay}b; see also 
\figref{fig:gammaDelta_mr_decay}b).\footnote{In \figref{fig:mr_decay}, 
the anisotropy appears to stabilise considerably above the marginal level 
in Stage 2 (secular growth), but in fact, if we continued the driven stage 
longer than one shear time, it would slowly decay towards the marginal level 
\citep{Kunz14}. Note further that the fast relaxation to 0 that occurs immediately 
after the shear is switched off occurs on a time scale that is consistent with 
the effective collision rate at the moment of switch-off: from \figref{fig:mr_decay}a,
$S/\nueff\sim0.02$. The subsequent approach to the mirror-marginal level is from 
the stable side ($\Delta\to1/\beta$ from below). The latter behaviour is analogous 
to what happens in the case of decaying firehose turbulence (see footnote~\ref{fn:fh_decay}).} 

The comparison of the effective collisionality with the scalings 
\exref{eq:nueff_mr_decay_trapped} and \exref{eq:nueff_mr_decay_passing}
is shown in \figref{fig:nueff_mr_decay}. The overall scaling should be 
closer to $\nueff\propto|\dBpar/B_0|^2$, because it is the weighted average 
\exref{eq:nueff_mr_decay}, but, unfortunately, 
the quality (and/or the degree of asymptoticity) of the data does not allow 
for a clear differentiation between the two scalings. 

This, with some caveats, our numerical results appear to be largely consistent with 
the simple description of the decaying mirror turbulence proposed above. 

\section{Pressure anisotropy and microturbulence in a reversing shear flow}
\label{sec:shear}

Let us now investigate the more ``general'' scenario in which 
the shear $S$ acts for a while---typically for a time of order $S^{-1}$---and 
then reverses, with the new, equal and opposite,
shear acting for another time of order $S^{-1}$, again reversing, etc. 
(in general, the shear will also 
change magnitude, but we will not study such a situation here).
Arranging such a sequence is a small step toward a local model of 
a macroscopic turbulent flow.

\subsection{From firehose to mirror}
\label{sec:fh_to_mr}

Consider first the case in which 
the shear switches from positive (firehose-unstable) to negative (mirror-unstable). 
As the mean field starts growing, a system with no initial microscale fluctuations 
would develop a positive pressure anisotropy and go mirror-unstable, as it did 
in the sheared initial phase of the numerical experiment shown in \figref{fig:mr_decay}. 
In contrast, in the numerical experiment we are about to describe here, 
the shearing of the mean field to larger values happens 
against the background of pre-existing firehose fluctuations, 
which will in general evolve in time and thus contribute to the 
evolution of the pressure anisotropy. 

This evolution satisfies \eqref{eq:Delta_fh}, 
except the direction of the shear is now reversed: 
\beq
\frac{1}{3}\frac{\rmd\Delta}{\rmd t} + \nueff\Delta 
= \lt|\frac{\rmd\ln B_0}{\rmd t}\rt| + \frac{1}{2}\frac{\rmd\bsq}{\rmd t}.
\label{eq:Delta_fh_to_mr}
\eeq
Note that at least initially, there are no mirror fluctuations (no $\dBpar$) because 
at the moment of shear reversal the system is at the firehose threshold 
and so is mirror-stable. 

There are two possible scenarios, depending on the magnitude of $\beta$. 

\subsubsection{Theoretical expectations: case of $\beta\ll\Omega/S$}
\label{sec:low_beta}

In this case, the firehose fluctuations at $St\sim1$ are saturated 
at $\bsq\sim(\beta S/\Omega)^{1/2}$ (\secref{sec:fh_growth_num}). 
As we found in \secref{sec:fh_decay}, in the absence of an externally imposed 
shear, they will decay at the rate $\gamma\sim\Omega/\beta$. 
It is reasonable to expect that their evolution in the presence of 
a shear that does not drive them will be no different provided this shear 
is slow compared to their free decay. Namely, 
the second term in the right-hand side of \eqref{eq:Delta_fh_to_mr} 
is larger than the first provided
\beq
\frac{\rmd\bsq}{\rmd t} \sim \gamma\bsq 
\gg \lt|\frac{\rmd\ln B_0}{\rmd t}\rt|\sim |S|.
\label{eq:cond_free_fh}
\eeq
With $\bsq\sim(\beta |S|/\Omega)^{1/2}$ initially, 
this is the case in precisely the regime that we are considering, 
$\beta\ll\Omega/S$. The firehose fluctuations will then decay freely 
according to 
\beq
\bsq\sim\lt(\frac{\beta |S|}{\Omega}\rt)^{1/2}e^{-\gamma t},
\eeq
until $\bsq$ becomes small enough for the condition \exref{eq:cond_free_fh}
to stop being satisfied. This happens at the time $t_\mathrm{decay}$ when 
\beq
\bsq\sim\frac{|S|}{\gamma}\sim\frac{\beta |S|}{\Omega}
\quad\Rightarrow\quad
t_\mathrm{decay} \sim \frac{\beta}{\Omega}\ln\frac{\Omega}{\beta |S|},
\eeq 
i.e., generally in just a fraction of one shear time. 
After this, the remaining firehose fluctuations can simply be 
sheared away (on the time scale $|S|t\sim\bsq\sim\beta |S|/\Omega\ll1$), 
the pressure anisotropy is no longer pushed towards the firehose-marginal 
level by their free decay, but instead pushed up by the shear 
towards the mirror threshold, and the system can go mirror unstable 
as it does in the case of no initial fluctuations, with the mirror-instability  
evolution proceeding as before (see \secref{sec:mr_growth}). 

\subsubsection{Theoretical expectations: case of $\beta\gg\Omega/S$}
\label{sec:high_beta}

At such ultra-high betas, the firehose fluctuations at $St\sim1$ are $\bsq\sim1$
and, once the shear is reversed, their free decay would be too slow to compete 
with the new shear (i.e., the condition \exref{eq:cond_free_fh} is broken from 
the outset). The dominant balance in \eqref{eq:Delta_fh_to_mr} is then between 
the two terms in the right-hand side: 
\beq
\frac{1}{2}\frac{\rmd\bsq}{\rmd t} \approx 
-\lt|\frac{\rmd\ln B_0}{\rmd t}\rt|\sim -|S|.
\eeq
Thus, the firehose fluctuations are sheared 
away secularly, 
\beq
\bsq\sim 1 - St, 
\label{eq:fh_shear_away}
\eeq
and will be gone after approximately one shear time. 
This means that during (most of) the time that the reversed shear acts, {\em the decay 
of the residual firehose fluctuations can offset the increase of the mean field 
by shear---and so the system need not cross the mirror threshold and go unstable.} 
By the time the residual firehose reservoir is exhausted and the system is ready 
for mirror instability, it will be too late as the shear will, in general, reverse again. 

\begin{figure*}
\begin{tabular}{cc}
\includegraphics[width=0.5\textwidth]{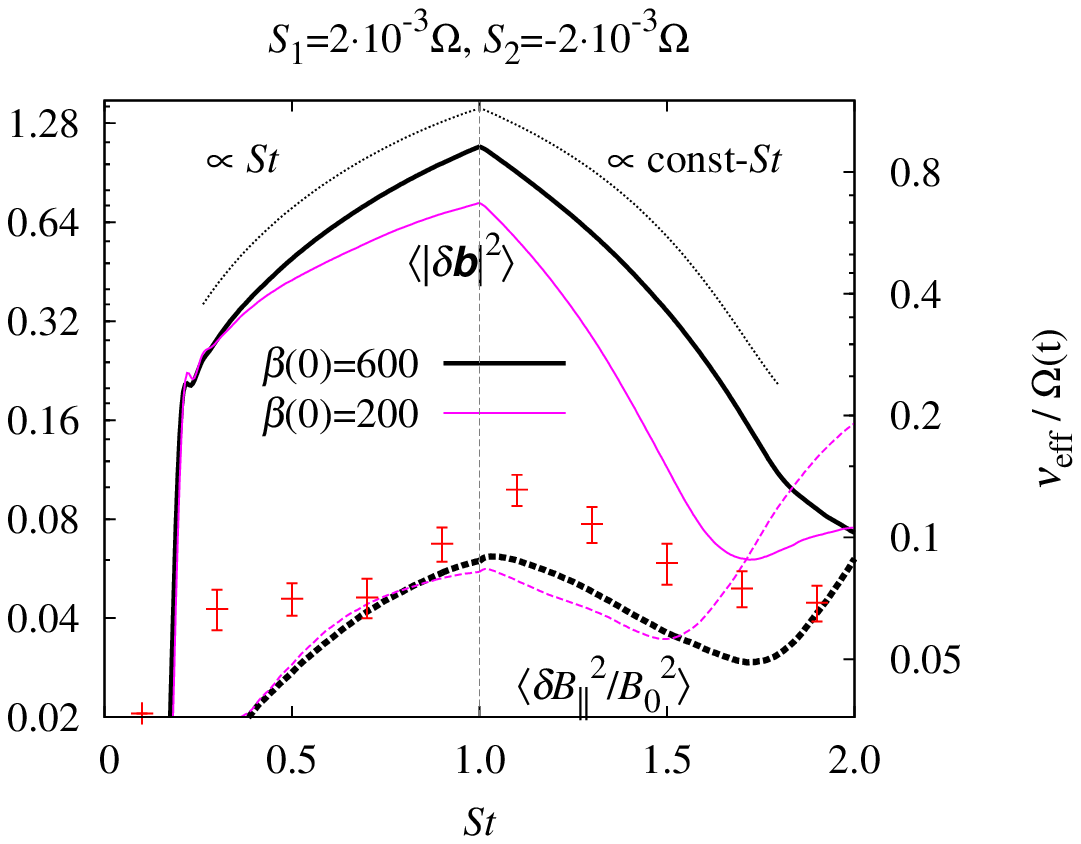}&
\includegraphics[width=0.5\textwidth]{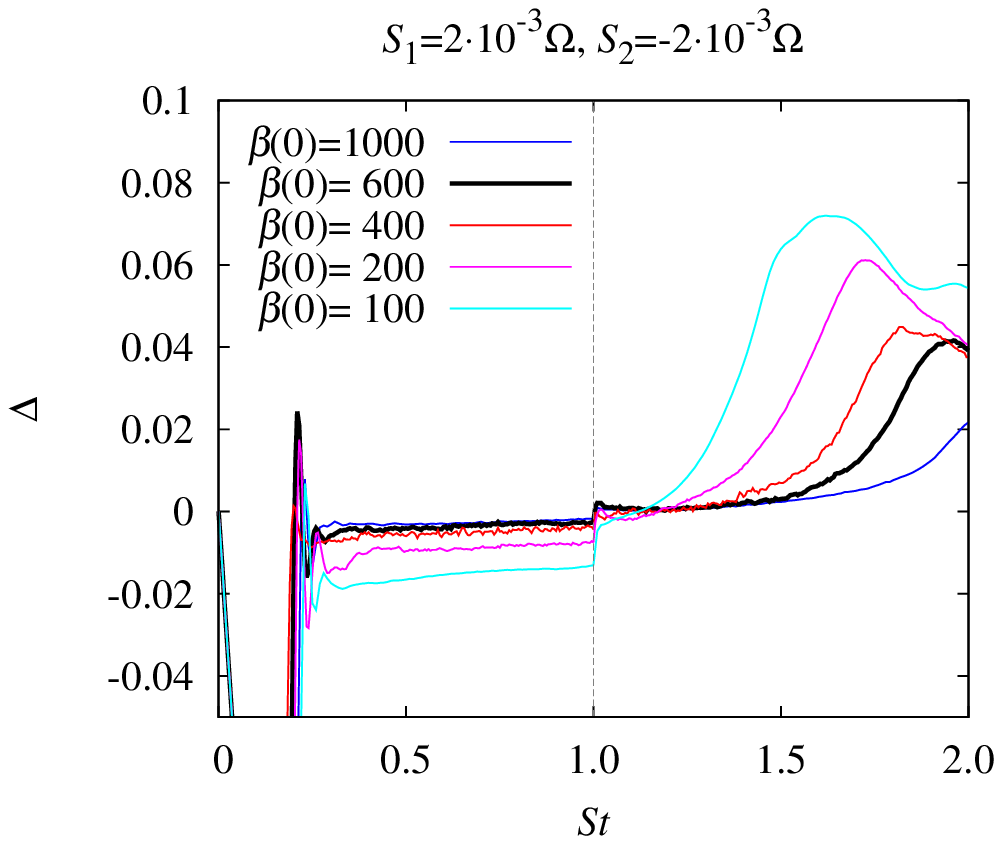}\\
(a)&(b)\\
\includegraphics[width=0.5\textwidth]{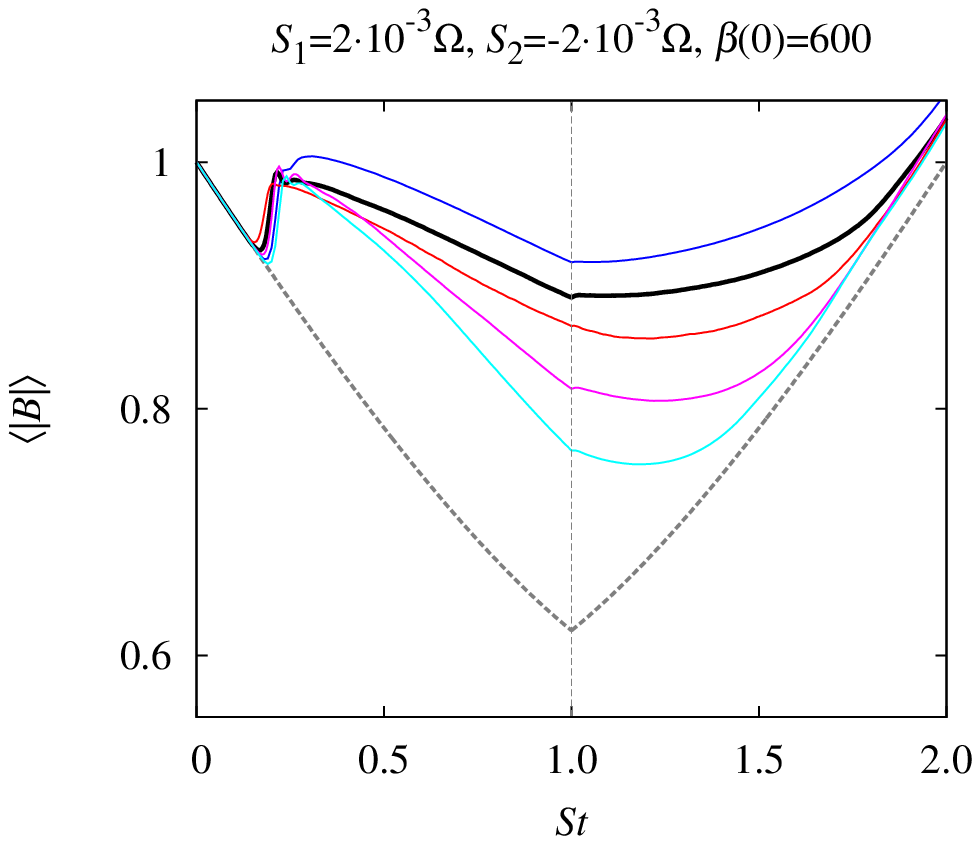}&
\includegraphics[width=0.5\textwidth]{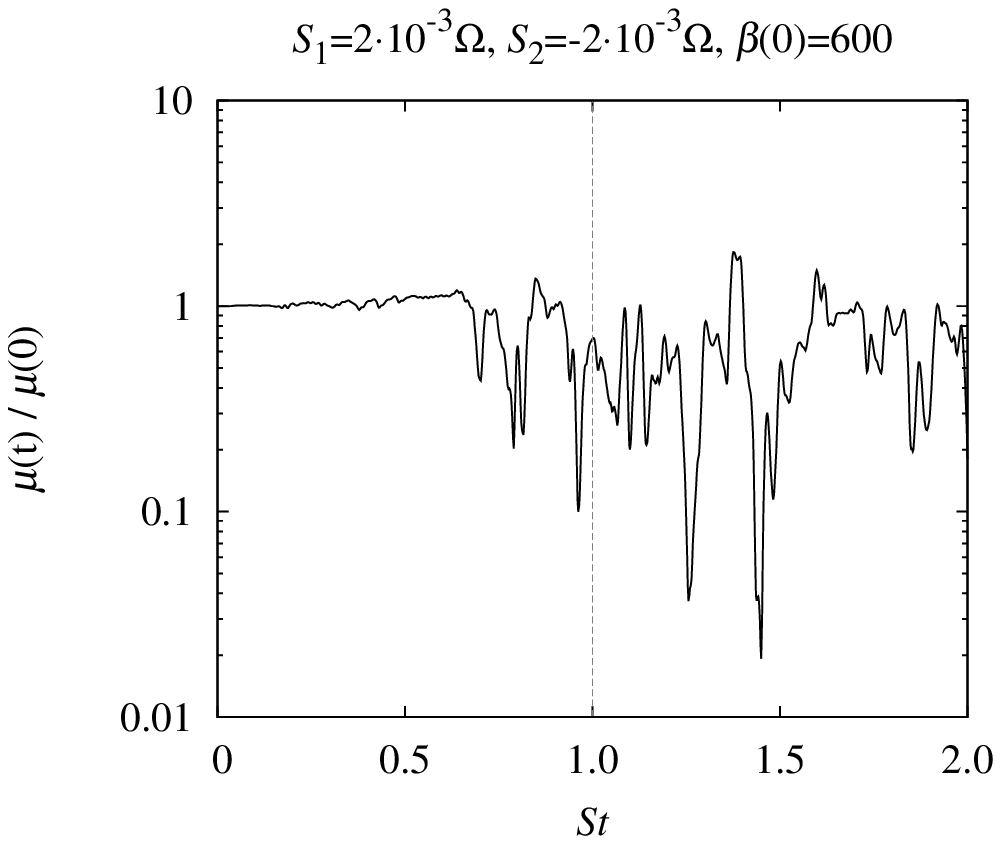}\\
(c)&(d)
\end{tabular}
\caption{Same as \figref{fig:fh_decay}, but with reversed shear in the 
second phase (driving firehose, then mirror): from 
$S=2\times10^{-3}\Omega$ until $St=1$ to $S=-2\times10^{-3}\Omega$ afterwards; 
$\beta(0)=600$. Same quantities are shown, with the following differences. 
(a) Time evolution of both the mirror perturbations, $\la\dBpar^2/B_0^2\ra$ 
(dotted lines), and the firehose ones, $\bsq$ (solid lines), is given
for two values $\beta(0)=600$ (black, bold) and $\beta(0)=200$ (magenta), 
showing that the onset of the mirror-unstable regime 
is delayed at higher $\beta$; the effective collisionality $\nueff$ is shown 
for the $\beta(0)=600$ case. 
(b) Time evolution of the pressure anisotropy $\Delta=(\pperp-\ppar)/p$ 
is given for several values of $\beta(0)$. 
At $St<1$, each of these simulations hovers near its respective firehose 
threshold, $\Delta=-2/\beta(t)$, similarly to \figref{fig:fh_decay}b 
(thresholds not shown here to avoid clutter).
At $St>1$, $\Delta$ starts growing towards the mirror threshold, 
but that happens later at higher $\beta$. 
(c) Time evolution of the mean magnitude of the field, 
$\la B\ra = \la|\vB|\ra$ (as opposed to the magnitude of the mean field, $B_0=|\la\vB\ra|$)
for the same values of $\beta(0)$ as in (b) (where the legend shows which line corresponds 
to which $\beta(0)$), 
showing that at higher $\beta$, $\la B\ra$ barely changes throughout both the firehose 
and mirror regimes. 
The magnitude of the mean field, $|\la\vB\ra|$ is shown for the case $\beta(0)=600$.}
\label{fig:fh_to_mr}
\end{figure*}

\subsubsection{Numerical results}
\label{sec:fh_to_mr_num}

The scenario proposed above is broadly confirmed by the numerical experiments with 
reversing the shear (the analogue of the one described in \secref{sec:fh_decay_num}, 
but the shear equal to $-S$, rather than 0, for $|S|t>1$). 
This is documented in \figref{fig:fh_to_mr}. 
The two values of beta, $\beta=200$ and $600$, 
for which the evolution of the fluctuation level is shown in \figref{fig:fh_to_mr}a 
correspond approximately 
(although not asymptotically) to the regimes discussed in \secsand{sec:low_beta}{sec:high_beta}, 
respectively. We see that the onset of the mirror-unstable regime 
(manifested by the growth of $\la\dBpar^2\ra$ in \figref{fig:fh_to_mr}a and 
of the pressure anisotropy $\Delta$ in \figref{fig:fh_to_mr}b) 
is delayed in both cases---and for a longer time 
at higher $\beta$. Runs with other values of $\beta$ confirm this trend 
(see \figref{fig:fh_to_mr}b) 
and in particular the assertion in \secref{sec:high_beta} that at high enough $\beta$, 
the onset of the mirror instability is delayed by approximately one shear time. 

Note that, in the ultra-high-beta regime, while the mean field decays and then grows again, 
the mean magnitude of the magnetic field (i.e., the total magnetic energy, including 
firehose fluctuations) stays essentially constant (see \figref{fig:fh_to_mr}c): in the 
driven-firehose regime ($|S|t<1$, $S>0$), this is because the stage of the secular
firehose growth compensating the decrease of the mean field extends over an entire 
shear time (see \secref{sec:fh_sat}), whereas in the decaying-firehose 
regime ($|S|t>1$, $S<0$), the fluctuations' decay by being sheared away is compensated 
by the growth of the mean field (\secref{sec:high_beta}).   

Snapshots of the firehose field being sheared away are shown in 
\figsref{fig:fh_pics}d and e ($St=1.4$ and $St=2$, respectively); 
while \figref{fig:fh_pics}f shows the (rather feeble) mirror field 
that comes to replace the firehose turbulence. 

\begin{figure*}
\begin{tabular}{cc}
\includegraphics[width=0.5\textwidth]{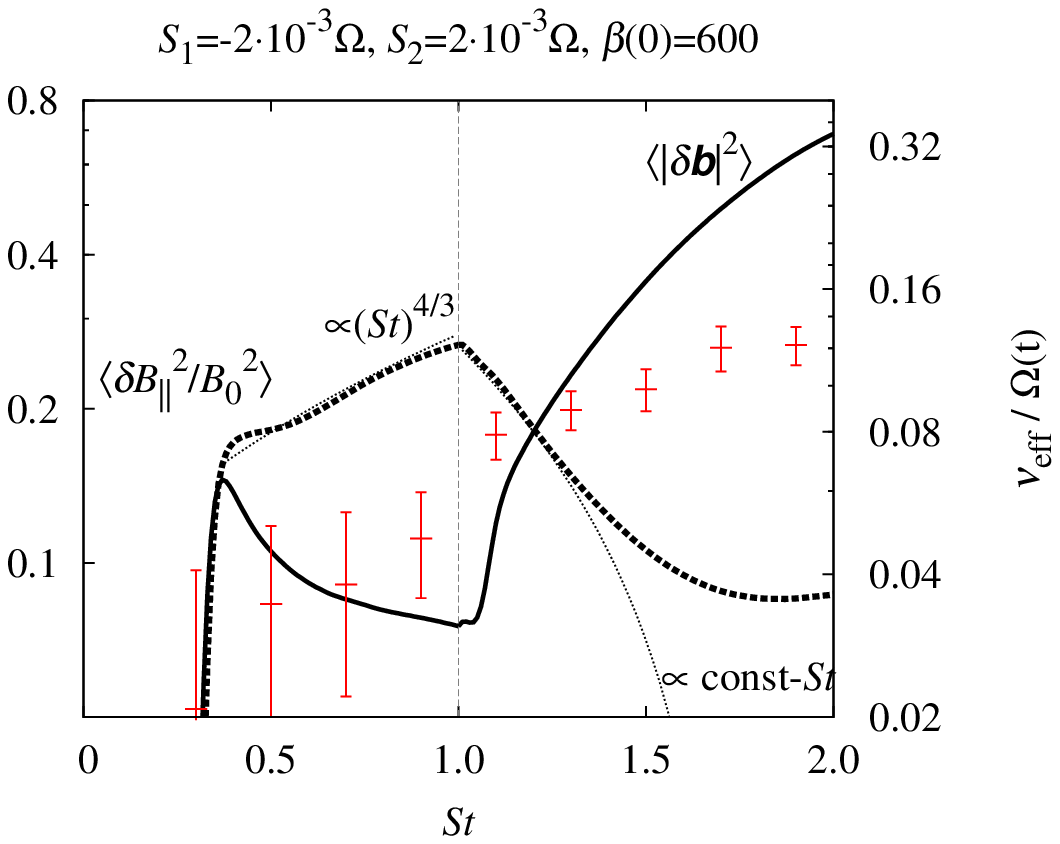}&
\includegraphics[width=0.5\textwidth]{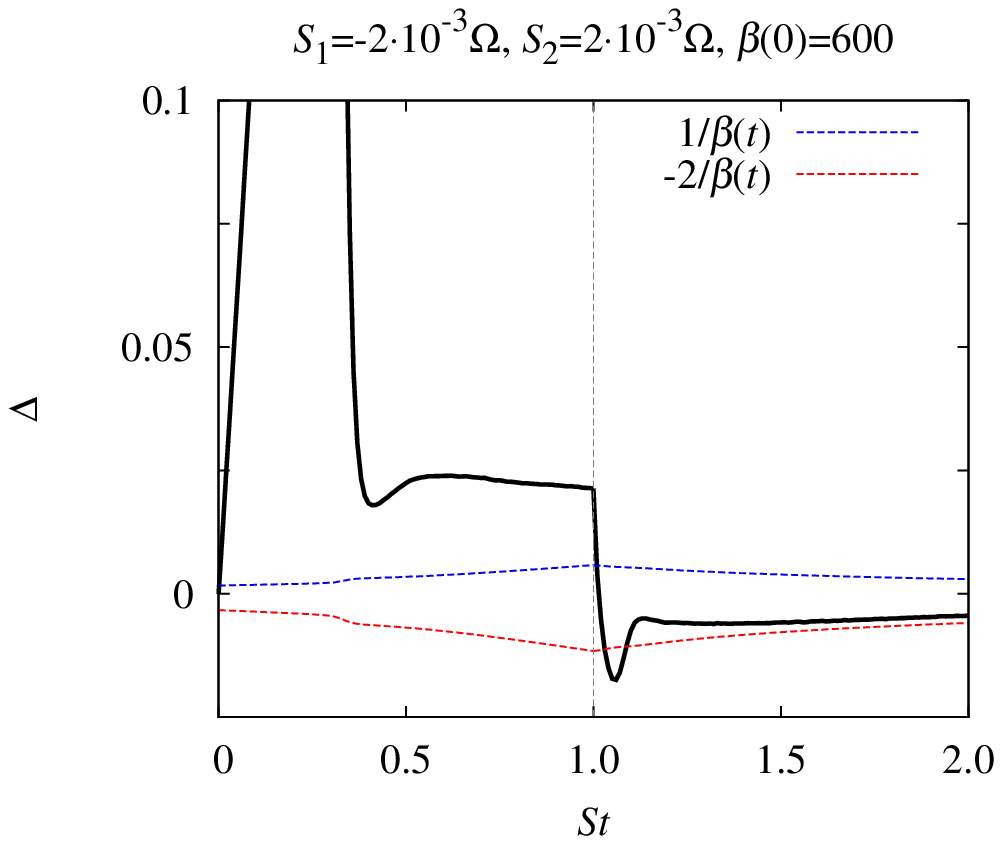}\\
(a)&(b)\\
\includegraphics[width=0.5\textwidth]{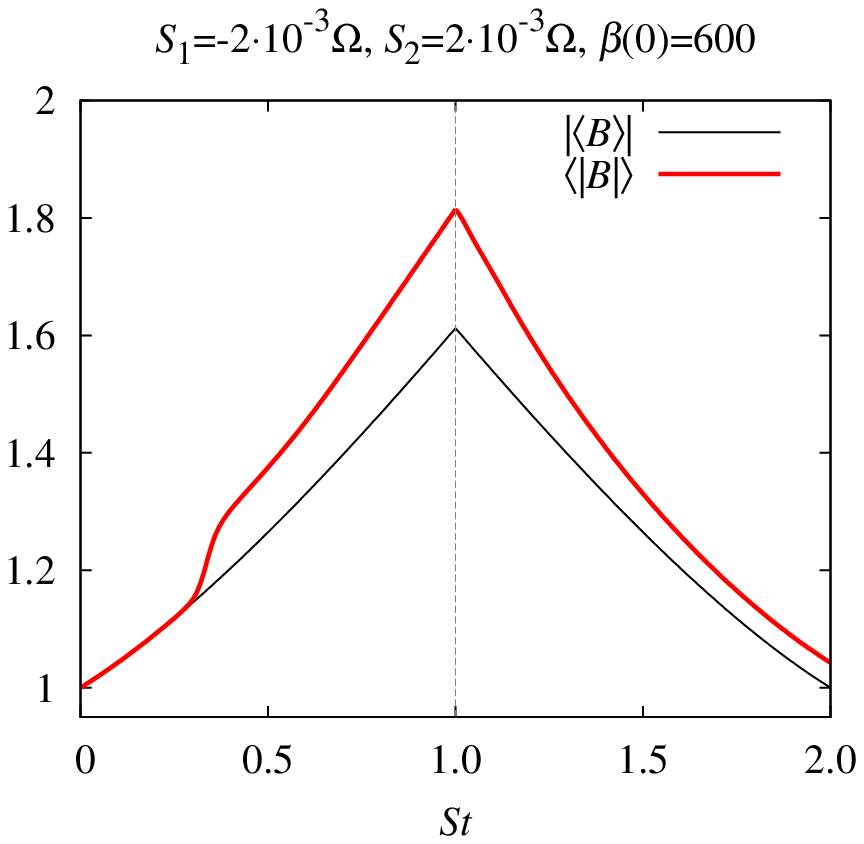}&
\includegraphics[width=0.5\textwidth]{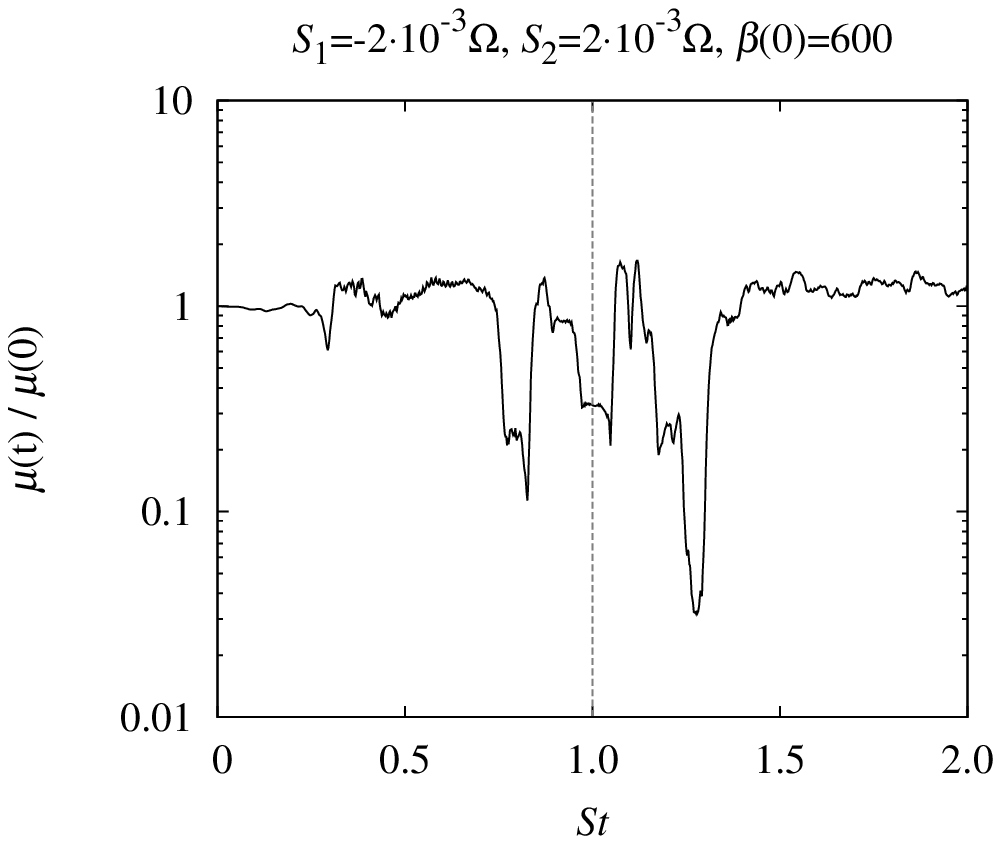}\\
(c)&(d)
\end{tabular}
\caption{Same as \figref{fig:mr_decay}, but with reversed shear in the 
second phase (driving first mirror, then firehose): from 
$S=-2\times10^{-3}\Omega$ until $|S|t=1$ to $S=2\times10^{-3}\Omega$ afterwards; 
$\beta(0)=600$. Same quantities are shown with the following differences. 
(a) Time evolution of both the mirror perturbations ($\la\dBpar^2/B_0^2\ra$, bold dotted line)
and the firehose ones ($\bsq$, bold solid line) is shown. 
Dotted lines show the theoretical estimates for the secular 
growth of driven mirror turbulence at $|S|t<1$, \eqref{eq:mr_secular}, 
and for its shearing away at $|S|t>1$, analogously to \eqref{eq:fh_shear_away}. 
Note the close similarity in the 
evolution of the system at $|S|t>1$ with the growing firehose turbulence 
that started from a pristine state---see the $|S|t<1$ part of 
\figsand{fig:fh_decay}{fig:fh_to_mr}. 
(b) Both the firehose and mirror thresholds are shown 
(red and blue dotted lines, respectively).}
\label{fig:mr_to_fh}
\end{figure*}

\subsection{From mirror to firehose}
\label{sec:mr_to_fh}

Now consider the opposite situation to \secref{sec:fh_to_mr}, 
{\em viz.}, a switch from negative (mirror-unstable) shear 
to positive (firehose-unstable) one. 
The results of this numerical experiment, illustrated by \figref{fig:mr_to_fh}, 
are perhaps not as newsworthy as those of the previous one 
(\secref{sec:fh_to_mr}). In summary, after the shear is reversed, 
mirror fluctuations decay, the system is driven through the firehose 
threshold and firehose fluctuations grow to keep it marginal. 
The mirrors decay by being sheared away, at least at 
$\beta\gtrsim\Omega/S$ (see \figref{fig:mr_to_fh}a). At $\beta\ll\Omega/S$, presumably 
their free decay would be faster (\secref{sec:mr_decay}) and would override the shearing, 
but this regime was not achieved in a pure form at the values $\beta\ge100$ that 
we have investigated---presumably a clean asymptotic separation between the shear 
and the free-decay time scales could be achieved in simulations with much smaller~$S$. 

The emergence of the firehose turbulence seems to occur without any delay and 
quite independently of the fact that there is an initial sea of mirrors, 
essentially in the same way as it did in \secref{sec:fh_sat}. 
This might appear to be a somewhat puzzling result in view of our experience 
with the system's travails (unsuccessful at ultra-high $\beta$) in switching from 
the firehose to the mirror regime in \secref{sec:fh_to_mr} and with the free decay 
of the mirror turbulence in \secref{sec:mr_decay}, where we argued that 
the decay of the mirrors could increase mean pressure anisotropy in the 
system---this begs the question of why then it does not impede or at least 
slow down the system's journey towards the firehose threshold.  

Without claiming to possess a complete theory (any more than we did in the preceding 
sections), we offer the following considerations to make sense of the observed 
behaviour. An important difference between mirror and firehose perturbations 
is the scales at which they develop: whereas the latter grow at 
$\kperp\rho_i\sim\kpar\rho_i\sim1$ \citep{Kunz14}, the former are generally larger 
and separated from the Larmor scale by the marginality parameter: 
$(\kperp\rho_i)^2 \sim \kpar\rho_i\sim \beta\Delta - 1$ \citep{Hellinger07}.  
It is indeed because of this relatively long-scale nature of the mirrors 
that they fail to scatter particles in their driven secular-growth regime.  
When the shear is reversed and the mirrors start decaying, one can think 
of the relatively long stretches of weaker or stronger field associated with 
the residual $\dBpar$ perturbations as ``local mean fields'' being sheared 
away, producing local patches of negative pressure anisotropy. 
These go firehose-unstable, producing Larmor-scale firehose fluctuations 
on top of the larger-scale, decaying mirror fields. 
The fact that the latter are somewhat inhomogeneous on the system scale 
does not bother the firehose fluctuations because of the smallness of their scale. 
The effect of the released trapped particles sampling larger field and thus increasing the 
anisotropy, as argued in \secref{sec:mr_decay}, appears to be overwhelmed 
in this experiment by the countervailing effect of the external firehose-driving 
shear (\figref{fig:mr_to_fh}b). 

Snapshots of both the mirror fluctuation field being sheared away and the 
firehose field emerging in parallel are shown in \figsref{fig:mr_pics}c and d, 
respectively (both at $St=1.4$). 

\begin{figure*}
\begin{center}
\includegraphics[width=\textwidth]{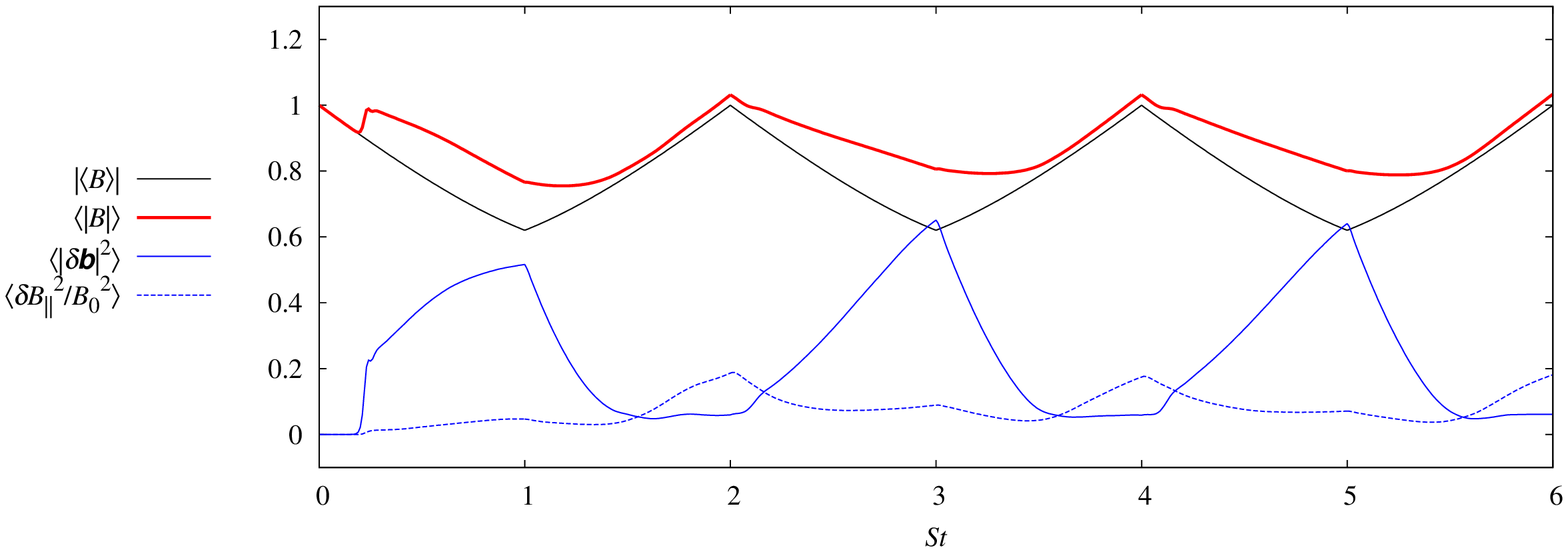}\\
(a) $\beta=100 < \Omega/|S|$, $|S|=2\cdot10^{-3}\Omega$\\
\includegraphics[width=\textwidth]{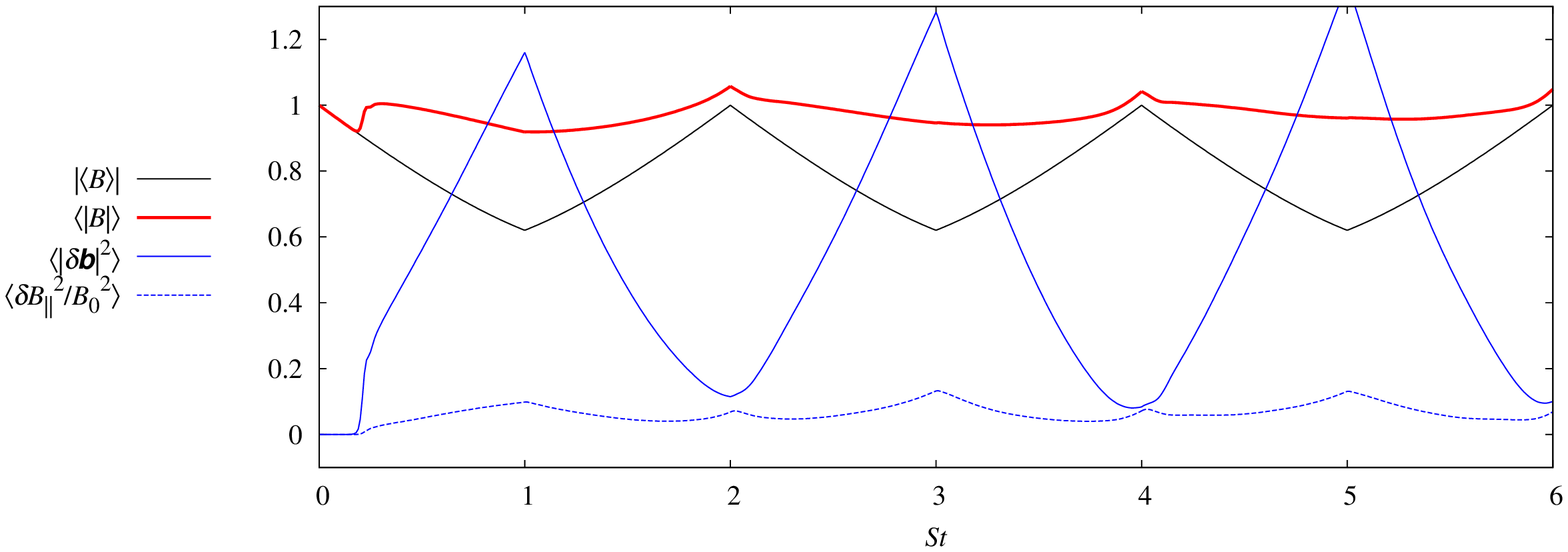}\\
(b) $\beta=1000 > \Omega/|S|$, $|S|=2\cdot10^{-3}\Omega$\\
\end{center}
\caption{Time evolution of the magnitude of the mean field $\la|\vB|\ra$ (solid black), 
mean magnitude of the field $\la B\ra$ (bold red), 
firehose fluctuation level $\bsq$ (solid blue) 
and mirror fluctuation level $\la\dBpar^2/B_0^2\ra$ (dashed blue) 
in two runs where the sign of shear was alternated several times: 
(a) the case of $\beta < \Omega/|S|$, (b) the case of $\beta > \Omega/|S|$. 
Note the much-delayed onset (amounting to near-complete suppression) of mirror 
instability and near constancy of $\la B\ra$ in the ultra-high-$\beta$ case. 
These plots should be compared to \figsand{fig:fh_to_mr}{fig:mr_to_fh}(a,c).} 
\label{fig:alt}
\end{figure*}

\section{Conclusions}
\label{sec:conc}

\subsection{Summary}
\label{sec:sum}

This paper is a contribution to the rapidly expanding body of evidence 
about the nature and behaviour of pressure-anisotropy-driven ion-scale 
instabilities \citep{McKean93,Quest96,Gary97,Gary98,Hellinger01,Matteini06,Kunz14,Riquelme15,Sironi15,Hellinger15,Hellinger15a}. 
These studies' {\em raison d'\^etre} is to understand the basic
ingredients that combine to provide the microphysical 
(microscale) background to the large-scale dynamics, hence the recent emphasis 
on driven systems where uniform shear \citep{Kunz14,Riquelme15} 
or compression/rarefaction \citep{Matteini06,Sironi15,Hellinger15,Hellinger15a} 
provide the energy source for the instabilities. The new step that we have 
taken here is to investigate what happens when the macroscale flow 
{\em changes} 
and a new macroscale configuration is superimposed on the microscale 
state left behind by the previous one. We have studied two basic cases: 
the free decay of microscale fluctuations when the drive---externally imposed 
shear---is removed and the effect of reversing the shear. We have discovered 
several interesting and, in our view, quite fundamental phenomena. 

\begin{itemize}

\item The free decay of both firehose (\secref{sec:fh_decay})
and mirror (\secref{sec:mr_decay}) fluctuations is constrained 
by the same marginal-stability conditions as their growth 
and saturation in the driven regime. When firehoses decay, 
magnetic energy decreases and so pressure anisotropy is pushed in the 
negative direction, towards the firehose threshold. To avoid crossing it, 
the fluctuations scatter particles and decay exponentially 
at the rate $\gamma\sim\Omega/\beta$. When mirrors decay, trapped 
particles are released, sample stronger magnetic fields and the 
pressure anisotropy is pushed in the positive direction, towards 
the mirror threshold. Again, the system avoids crossing this threshold 
because fluctuations scatter particles (for mirrors, this is unlike what they 
do in the driven case; see \secref{sec:mr_growth}) 
and decay on the same $\beta$-dependent time scale as the firehose ones,
although the decay law is slower than exponential (\eqref{eq:mr_decay}).\\ 

\item When the shear is reversed, rather than switched off, the free 
decay described above competes with fluctuations of one type (mirror 
or firehose, depending on the sign of the old shear) being 
sheared away and fluctuations of the other type being driven. 
A threshold value of $\beta$ emerges: when $\beta\ll\Omega/S$, 
the decay rate is large, $\gamma\gg S$, so the free decay is fast compared 
to macroscale dynamics, old fluctuations quickly disappear and new 
ones can be made from scratch; in contrast, when $\beta\gtrsim\Omega/S$ (``ultra-high'' beta), 
the old microscale state is only removed on the macroscopic (shear) timescale. 
The consequences of this are asymmetric for the firehoses (\secref{sec:fh_to_mr})
and mirrors (\secref{sec:mr_to_fh}): 
the shearing away of residual firehose turbulence prevents 
growth of the pressure anisotropy and thus stops the system from 
going mirror-unstable; the shearing away of a residual field of mirrors 
does not stop the system from going firehose-unstable and producing copious 
firehose fluctuations. We conjectured that the reason for the latter is 
the relatively large scale of the mirrors, which are ``seen'' by the 
Larmor-scale firehoses essentially as local instances of a decreasing 
mean field.\\ 

\item By way of illustration of the key features of the moderate- 
and ultra-high-beta regimes, we refer the reader to \figref{fig:alt}. 
This shows two numerical experiments, one with $\beta<\Omega/S$, the other 
with $\beta > \Omega/S$, in which the sign of the shear is alternated several 
times, so the system goes through multiple successive periods of decaying (firehose-unstable) 
and growing (mirror-unstable) mean field (this is analogous to successive 
instances of the experiments described in \secsand{sec:fh_to_mr}{sec:mr_to_fh}).
The key point is that at ultra-high $\beta$, periods of firehose growth 
and decay alternate, the system spends most of its evolution close to the 
firehose threshold, the mirror instability is almost completely suppressed,  
while the mean magnitude of the magnetic field barely changes at all; in contrast, 
at moderate $\beta$, the periods of decaying/growing mean field feature 
(relatively) lively firehose/mirror fluctuations which however do not survive long 
past the point when they cease to be driven by the external shear.\\ 

\item In the context of the emergence of the threshold value of $\beta$, 
we have also carried out a $\beta$ scan of the driven firehose turbulence, 
to complement a previous study by \citet{Kunz14}, where only one 
value of $\beta$ was used. It has turned out (\secref{sec:fh_sat}) that 
the saturation level of firehose fluctuations 
depends on $\beta$, {\em viz.}, $\bsq\sim (\beta S/\Omega)^{1/2}$, so when 
$\beta\gtrsim\Omega/S$, we have $\bsq\sim1$. The latter saturation regime 
takes the system on the order of one shear time to achieve and follows after 
a long phase of secular growth, during which the pressure anisotropy 
is pinned at marginal stability purely by the increase in the fluctuation energy, 
with no appreciable scattering of particles. This is broadly in line 
with the earlier theoretical argument of \citet{Sch08} and \citet{Rosin11}, although 
unlike in their study, the fluctuations are oblique, rather than parallel. 
We do not as yet have a complete theoretical understanding of this regime, 
or of the $\beta$ and $S$ dependence of the saturated firehose amplitude. 

\end{itemize}

\subsection{Implications for cosmic magnetogenesis} 
\label{sec:mag}

As large-scale dynamical motions that drive pressure anisotropies are 
invariably associated with changing the macroscopic magnetic field 
and since the resulting instabilities give rise to microscopic magnetic 
fluctuations, the implications of all this for the long-standing 
problem of the origin (and maintenance) of cosmic magnetism are 
the obviously relevant and interesting topic to discuss. 

\subsubsection{Plasma dynamo and ICM viscosity}
\label{sec:dynamo}

Let us consider what our results might imply for the operation of 
``plasma dynamo''---the process of amplification of magnetic fields 
by turbulence in the ICM or a similar plasma that is too weakly 
collisional for an MHD description to be legitimate. 

As far as a dynamically 
weak magnetic field is concerned, macroscale turbulence is a random sequence 
of rates of strain (stretchings, shears, compressions, etc.---we shall 
call them all shears for brevity), changing 
on a macroscopic time scale (``eddy-turnover time'') and leading 
to a random sequence of increases and decreases in the field strength 
(magnetic energy). In standard MHD, in 3D, this results on average 
in an exponential growth of the magnetic energy at the rate 
comparable to the rms rate of strain of the velocity field 
\citep{Moffatt64,Zeldovich84,Ott98,Chertkov99,Sch04,MHDbook}. 

As far as the microphysical world is concerned, a sequence 
of random shears leads to a sequence of instances of positive or negative 
pressure anisotropies exciting mirror or firehose fluctuations (which 
is indeed what is seen in direct numerical simulations of collisionless 
dynamo; see \citealt{Rincon16}). 
The magnetic energy generated in the form of these fluctuations and/or 
the anomalous particle scattering caused by them lead to a 
modification of the magnetic field's evolution in time. 
Various scenarios for this were analysed by \citet{Mogavero14}, focusing 
especially on two opposing possibilities: growth/decay of the magnetic field 
caused by the macroscopic plasma motion being slowed down by microscale 
energy changing to cancel the macroscale evolution vs.\ anomalous particle 
scattering causing a decrease in the effective viscosity of the plasma
(because dynamic viscosity is $\sim p/\nueff$), 
thereby allowing the turbulent motions to reach smaller scales and hence 
develop greater rates of strain and amplify the field faster. 

We are now in a position to amend this analysis in light of new 
information provided by \citet{Kunz14}, \citet{Riquelme15}, \citet{Rincon15} 
and the present study. 

\subsubsection{Plasma dynamo at moderate $\beta$}
\label{sec:dynamo_mod}

As was shown analytically by \citet{Rincon15} and numerically by \citet{Kunz14} 
(see review in \secref{sec:mr_growth} of this paper), driven microscale mirrors 
can keep the plasma at the mirror threshold without scattering particles 
for as long as a shear time. They do this by sequestering ever more particles 
in increasingly deep mirror wells so as to prevent the plasma on average 
from ``seeing'' that the mean field is increasing. There is no net decrease 
of magnetic energy to cancel the mean-field growth and so the mirror 
instability does not change the local rate of growth of the field, 
so long as the ``eddies'' that give rise to these growth episodes 
decorrelate on the same time scale as they ``turn over.''\footnote{Some level 
of scattering can occur if a growth episode lasts so long that the mirrors 
reach $\dB/B\sim1$ \citep{Kunz14,Riquelme15}. Then the local effective viscosity 
of the plasma drops and the ``eddy'' that caused the growth of the field breaks up into 
smaller ``eddies''---these can produce even faster growth, but also decorrelate 
very quickly. See the discussion that is about to follow
of similar behaviour in the firehose regime.} 
Thus, it is probably a good assumption that growth episodes 
during which plasma crosses the mirror threshold are typically associated with 
large-scale motions that stay correlated on approximately one shear time, 
produce local regions filled with mirror wells and do not 
scatter particles. 

In contrast, driven firehose fluctuations in the moderate-$\beta$ regime ($\beta\ll\Omega/S$)
do start scattering particles very quickly after the instability 
is triggered (see \citealt{Kunz14} and \secref{sec:fh_sat} 
of this paper) and so should correspondingly decrease the 
local viscosity of the plasma quasi-instantaneously compared to the 
time scale of the macroscopic motions.

Let us make some estimates in the modelling spirit of \citet{Mogavero14}. 
In the driven firehose regime (shear $S$ attempting to decrease mean field), 
the effective collisionality of the plasma is $\nueff\sim S\beta$ (\eqref{eq:fh_nueff}).
Then the effective dynamical viscosity is
\beq
\mueff\sim\frac{p}{\nueff}\sim\frac{B^2/8\pi}{S}. 
\label{eq:mu_mod}
\eeq  
Assuming velocities $U$ at the (macroscopic) outer scale $L$
and a Kolmogorov cascade supporting a constant flux of energy
\beq
\eps \sim \frac{\rho U^3}{L},
\eeq
where $\rho$ is the mass density of the plasma, 
the largest rate of strain will be at the cutoff scale set by the effective
viscosity \exref{eq:mu_mod}: by simple dimensional analysis,
\beq
S \sim \lt(\frac{\eps}{\mueff}\rt)^{1/2} 
\sim \Ma\lt(\frac{U}{L}\nueff\rt)^{1/2},
\label{eq:S_gen}
\eeq
where $\Ma=U/\vth$ is the Mach number ($\vth$ is the ions' thermal speed)
and we have used $p\sim\rho\vth^2$. 
Since $\nueff\sim S\beta$, we get 
\beq
S \sim \frac{U}{L}\,\Ma^2\beta \equiv \frac{U}{L}\,\Reff^{1/2}
\quad\Rightarrow\quad
\Reff = \Ma^4\beta^2,
\label{eq:S_mod}
\eeq
where $\Reff$ is the effective Reynolds number.\footnote{Note that 
the fact that $\Reff$ ``knows'' about the local value of $\beta$ and 
so the instantaneous strength of the magnetic field can set the size 
of the dominant rate of strain [\eqref{eq:S_vs_B}] means that plasma 
dynamo is \emph{never} strictly kinematic.}  
Another way to write the above estimate for the dominant shear~is 
\beq
S \sim \frac{\eps}{B^2/8\pi},
\label{eq:S_vs_B}
\eeq
i.e., weak magnetic fields are subject to faster change and the rate of this  
change is equal to the rate of power injection into the turbulence: 
\beq
\frac{\rmd}{\rmd t}\frac{B^2}{8\pi} \sim S\,\frac{B^2}{8\pi} \sim \eps. 
\eeq
This was the basis for a prediction by \citet{Mogavero14} that, in such 
an enhanced-collisionality plasma, magnetic energy can grow to equipartition 
with kinetic energy of the motions ($B^2/8\pi \sim \rho U^2/2$) in just one 
large-scale turnover time $\sim L/U$. However, we now know (\secref{sec:fh_decay})
that the collisionality-enhancing firehose fluctuations can only be maintained 
for a short time ($\sim \gamma^{-1}\sim\beta/\Omega\ll S^{-1}$) 
if they are not driven, as they indeed are not, in the instances of growing field. 
Therefore, the regions of high-$\Reff$ turbulence enabled by the firehose 
fluctuations are likely to break up and dissipate the macroscale field-decreasing 
motion that drove the firehoses in the first place more efficiently than they 
are to amplify the field. 

With the above estimates in hand, we can work out what it takes for 
the condition $\beta\ll\Omega/S$ to be satisfied: 
noticing that $\Omega=\vth/\rho_i = \vth/d_i\sqrt{\beta}$, where $d_i=c/\opi$ 
is the ion inertial scale, independent of the magnetic field, we get 
\beq
\beta \ll \frac{\Omega}{S}
\quad\Leftrightarrow\quad
\beta \ll \lt(\frac{L}{d_i}\frac{1}{\Ma^3}\rt)^{2/5}\equiv\betac.
\label{eq:betac}
\eeq  
It is not hard to check that the viscous cutoff scale for this 
turbulence, $\lvisc\sim L\Reff^{-3/4}$, is safely larger than $\rho_i$ 
at these values of $\beta$. 

Thus, assuming $\beta\ll\betac$, 
the moment a large-scale ``eddy'' somewhere tries to decrease the magnetic 
field, the plasma there explodes into a sea of firehoses, locally acquires 
a large effective Reynolds number as per \eqref{eq:S_mod},
and so the offending ``eddy'' quickly breaks up into smaller ``eddies,'' 
some of which might try to decrease the field further and so would 
break up into even smaller ones, and others to increase it, 
which would near-instantly cause them to regain viscosity and be damped by it
(the firehose fluctuations that were providing the increased 
collisionality quickly decay away in this regime; see \secref{sec:fh_decay}). 

To summarise, one might imagine the turbulent plasma dynamo in this regime 
as a sequence of random macroscale shears, 
some of which increase the macroscale magnetic field largely unimpeded 
but at the price of 
infesting it with microscale mirror ``bubbles,'' while others attempt to 
decrease it but instantly break up. Overall, this picture appears to favour 
magnetic-field amplification over its decay and thus should be a more efficient 
dynamo than its MHD counterpart---good news for fast cosmic magnetogenesis.   

We stress again that, because we have assumed $\beta\ll \Omega/S$, 
the firehose and mirror fluctuations decay quickly when the shear goes away or 
its direction changes (see \secsand{sec:fh_decay}{sec:mr_decay}) 
and so local patches of the system where these shears 
operate switch quasi-instantaneously from a firehose- to 
a mirror-dominated state, or vice versa.    

\subsubsection{Plasma dynamo at ultra-high $\beta$}
\label{sec:dynamo_ultra}
 
At ultra-high $\beta$, {\em viz.}, $\beta\gtrsim\Omega/S$, 
the main difference is that the system cannot quickly ``forget'' 
about its previous microscale state when the macroscale shear changes: 
namely, it has to rely on the new shear not only to amplify or diminish 
the macroscale magnetic field but also to shear away the microscale 
fluctuations left in the wake of the previous shear. 
By the time the old fluctuations are gone, the new shear 
decorrelates and changes again. 

As we saw in \secref{sec:fh_to_mr}, the shearing away of residual firehose 
fluctuations that occurs in the regions where the macroscale motion tries to amplify 
the field effectively prevents pressure anisotropy from growing enough to cross 
the mirror threshold. Therefore, in the ultra-high-$\beta$ regime, the system can largely 
avoid being mirror unstable and a given shear can increase 
magnetic field with impunity (as long as it does not last longer than 
a shear time and occurs after a firehose-unstable episode; \secref{sec:fh_to_mr}). 
However, as we saw in \secsand{sec:fh_decay}{sec:fh_to_mr}, the decaying 
firehose turbulence vigorously scatters particles.
According to \eqref{eq:nueff_decay}, it does so at the rate 
$\nueff\sim\Omega$ after the shear is reversed, at least for 
a large initial fraction of the shear time during which 
the residual firehose fluctuation level is still $\bsq\sim1$. The plasma 
in such a collisional regime is nearly unmagnetised. 
Using \eqref{eq:S_gen} with $\nueff\sim\Omega$, we get 
\beq
S \sim \frac{U}{L}\,\Reff^{1/2}, 
\quad
\Reff\sim \Ma\,\frac{L}{d_i}\frac{1}{\sqrt{\beta}}.
\eeq 
Note that it immediately follows that the plasma is in 
the ultra-high-$\beta$ regime provided
\beq
\beta\gtrsim \frac{\Omega}{S}
\quad\Leftrightarrow\quad
\beta\gtrsim\betac,
\label{eq:betac_ultra}
\eeq
where the threshold value $\betac$ is again given by \eqref{eq:betac}. 

The large shears generated in such high-$\Reff$ regions will, 
if they are locally field-decreasing, produce more firehose fluctuations
or, if they are field-increasing, shear them away and thus return 
to low collisionality and high viscosity, which will promptly damp away 
these shears. Again, somewhat analogously to what we saw happen 
in the driven-firehose regions analysed in \secref{sec:dynamo_mod}, 
the integrity of the macroscale motions is undermined by the firehose 
turbulence. 

Consider now the instances of decaying field against the background of residual 
mirror fluctuations. In \secref{sec:mr_to_fh}, we saw that these mirrors 
do not matter: firehose fluctuations are wont to grow on scales smaller 
than the mirrors and so the latter ``look'' like a local mean field to the former. 
As the driven firehose fluctuations in the ultra-high-$\beta$ regime grow 
secularly for about a shear time, rather than scattering particles (\secref{sec:fh_sat}), 
the decrease of the mean field in the firehose regime is accompanied by the compensating 
increase of the firehose turbulence and the total magnetic energy does not change. 

To summarise, in the ultra-high-$\beta$ regime, plasma motions attempting 
to decrease the mean field do not diminish the total magnetic energy because it 
is transferred to firehose fluctuations; the motions attempting to  
amplify the field, while prevented by those firehose fluctuations from 
rendering the plasma mirror-unstable, are quickly broken up because 
the fireshose turbulence scatters particles and thus radically diminishes
the effective viscosity of the plasma. It is hard to tell without a more 
quantitative theory how much field amplification these motions can accomplish 
before they are broken up and damped away. In the absence of such a theory, 
while it is not clear how quickly the field is amplified, it 
appears reasonably certain that decreasing the field is quite hard. 

 

 
\subsubsection{$\beta$ threshold}
\label{sec:beta_threshold}

One key conclusion of this work is that,  
since the rate of free decay of microscale fluctuations depends inversely 
on $\beta$ (\secsand{sec:fh_decay}{sec:mr_decay}), 
in ultra-high-$\beta$ plasmas it is quite difficult to get rid of magnetic 
energy once it has been generated. 
The threshold value of $\beta$ for this regime is $\betac\sim\Omega/S$, or, 
using $\Omega = \vth/d_i\sqrt{\beta}$, 
\beq
\beta_c \sim \lt(\frac{\vth}{S d_i}\rt)^{2/3} \sim \lt(\frac{L}{d_i}\frac{1}{\Ma}\rt)^{2/3}, 
\eeq
where the last expression was obtained via the estimate $S\sim U/L$, 
with $U$ and $L$ being the velocity and scale of the macroscopic motions that 
provide the local shear.
For standard core-ICM conditions \citep[see, e.g.,][]{Ensslin06,Sch06,Kunz11,Rosin11,Zhuravleva14}, 
taking $L\sim 10$~kpc $\sim10^{17}$~km, $d_i\sim10^3$~km, $\vth\sim10^3$~km/s, 
$\Ma \sim 0.1$, we find $\betac\sim10^{9}$, corresponding to an ICM magnetic 
field of $B\sim10^{-9}$~G. 

However, as we have seen in \secref{sec:dynamo_mod},  
the maximum rate of strain $S$ in the plasma flow may itself depend on $\beta$ 
because of the modification of the effective viscosity by anomalous particle 
scattering off microscale fluctuations. As a result, the estimate for the 
$\beta$ threshold is modified and given by \eqref{eq:betac} 
(see also \eqref{eq:betac_ultra}).
Using the same fiducial ICM parameters, 
this value is $\betac\sim10^{7}$, corresponding to $B\sim10^{-8}$~G. 
While this is weaker than the observed fields (which tend to be $B\sim10^{-6}$~G; 
see, e.g., \citealt{Carilli02}, \citealt{Govoni04}, \citealt{Vogt05}), 
it is substantially greater than even the most optimistic end of the range 
in which the primordial seed field is believed to lie \citep[$B\sim10^{-21}-10^{-9}$~G; see][]{Durrer13}.  

Thus, whichever estimate one uses, the ``ultra-high'' beta is in fact not all that high 
and both beta ranges considered in this paper should therefore be relevant to our understanding 
of cosmic magnetogenesis, with the dynamo in the ultra-high-beta regime 
possibly responsible for most of the magnetic-field amplification from 
the primordial seed.

\section*{Acknowledgments} 

We are grateful to S.~C.~Cowley for many important discussions, without which 
this work would not have been conceived. 
We also thank P.~Catto, F.~Parra, E.~Quataert, F.~Rincon, 
and A.~Spitkovsky for valuable comments. 
M.~W.~K.\ was supported by a Lyman Spitzer, Jr.\ Fellowship
and by the Max-Planck--Princeton Center for Plasma Physics. 
He thanks Merton College, Oxford, for its support of his visits to Oxford. 
All three authors also thank the Wolfgang Pauli Institute, Vienna, for its 
hospitality and support.

\bibliographystyle{mn2e}
\bibliography{msk_MNRAS}

\appendix

\section{Numerical set up}
\label{ap:code}

The PEGASUS code used for the numerical experiments reported here is described 
in detail in \citet{Kunz14jcp}. Here we provide a brief summary and some details 
relevant to this particular study.  

\subsection{Equations}

The code solves, by a standard $\delta f$-PIC method, the following equations. 
The ion distribution function $f_i$ satisfies
\begin{align}
\nonumber
&\lt(\frac{\dd}{\dd t} - Sx\,\frac{\dd}{\dd y}\rt)f_i 
+ \vv\cdot\vdel f_i\\
&\qquad+\lt[\frac{Ze}{m_i}\lt(\vE + \frac{\vv\times\vB}{c}\rt) + S v_x\vy\rt]
\cdot\frac{\dd f_i}{\dd\vv} = 0,
\end{align}
where $S$ is the externally imposed linear shear, $Ze$ and $m_i$ the ion charge
and mass, respectively. The magnetic field $\vB$ satisfies Faraday's equation
\beq
\lt(\frac{\dd}{\dd t} - Sx\,\frac{\dd}{\dd y}\rt)\vB = -c\vdel\times\vE - S B_x\vy. 
\eeq 
The electric field $\vE$ satisfies the standard plasma Ohm's law with isothermal electrons: 
\beq
\vE = -\frac{\vu_e\times\vB}{c} - \frac{T_e\vdel n_e}{e n_e},
\eeq 
where $T_e$ is the (constant) electron temperature, 
the electron flow velocity is computed from the ion flow velocity 
and the current (which is found via Amp\`ere's law),
\beq
\vu_e = \vu_i - \frac{\vj}{en_e} = 
\frac{1}{n_i}\int d^3\vv\,\vv f_i - \frac{c\vdel\times\vB}{4\pi e n_e},
\eeq
and the electron density is related to the ion densities by quasineutrality,
\beq
n_e = Z n_i = Z \int d^3\vv\,f_i.
\eeq
These equations are solved in a square box subject to shearing-periodic boundary conditions. 

\subsection{Code units and parameters}

The code units are: of time---the inverse initial ($t=0$) ion cyclotron frequency, 
\beq
[t] = \Omega^{-1}(0) = \frac{m_i c}{Ze B(0)},\quad
[S] = \Omega(0),
\eeq
of distance---the initial ion inertial length,
\beq
[x] = [y] = d_i(0) = \frac{c}{\omega_{{\rm p}i}(0)} = \frac{c}{Ze}\sqrt{\frac{m_i}{4\pi n_i(0)}},
\eeq
of velocity (therefore)---the initial Alfv\'en speed
\beq
[v] = d_i(0)\Omega(0) = \vA(0) = \frac{B(0)}{\sqrt{4\pi m_i n_i(0)}},
\eeq
of magnetic field---the initial magnetic field
\beq
[B] = B(0), 
\eeq
and of electric field
\beq
[E] = \frac{\vA(0) B(0)}{c};
\eeq
we also normalise particle number densities to their initial values,
\beq
[n_i] = n_i(0),\quad [n_e] = Zn_i(0).  
\eeq 

Under the above normalisations, 
each run is characterised by three physical dimensionless parameters: 
\beq
\frac{S}{\Omega(0)},\quad
\beta(0) = \frac{8\pi n_i(0) T_i(0)}{B^2(0)},\quad
\tau(0) = \frac{T_i(0)}{ZT_e},
\eeq
where $n_iT_i$ is the ion pressure. 
In all simulations reported here, we chose $\tau(0)=1$. 
The initial field was chosen to be 
\beq
\vB(0) = \frac{2\vx + 3\vy}{\sqrt{13}},
\eeq
so that the mean field was diagonal, $\la B_x\ra = \la B_y\ra$, at $St=1/2$. 

\subsection{Shear reversals}

The transformation of variables 
\beq
t \to - t,\quad
\vv \to -\vv,\quad
\vB \to -\vB
\eeq
is exactly equivalent to reversing the sign of shear
\beq
S \to -S.
\eeq
For a number of technical reasons, this proved a convenient ruse to 
implement the shear reversals for the numerical 
experiments described in \secref{sec:shear}. 

\subsection{Resolution}

All runs were done in square boxes whose 
size was $36\times36\,\rho_i^2(0)$, where $\rho_i(0)=\sqrt{\beta(0)}\,d_i(0)$. 
The cell (mesh) size was $0.125\rho_i(0)$, with 1024 particles per cell. 
The time step was $0.01\Omega^{-1}(0)$. 

The exception to this rule were the following runs 
done at a different resolution and/or in bigger boxes: 
\begin{itemize}
\item The $\beta=100$, $S=2\times10^{-3}\Omega$ run 
with alternating shear, shown in \figref{fig:alt}a: 
box size $576\times576\,d_i^2(0) = 57.6\times57.6\,\rho_i^2(0)$, 
cell size $1d_i(0)=0.1\rho_i(0)$ ($576^2$ cells), 
576 particles per cell.
\item The $\beta=1000$, $S=2\times10^{-3}\Omega$ run
with alternating shear, shown in \figref{fig:alt}b: 
box size $1152\times1152\,d_i^2(0) \approx 36.43\times36.43\,\rho_i^2(0)$, 
cell size $3d_i(0)\approx 0.095\rho_i(0)$ ($384^2$ cells), 
576 particles per cell.
\item The $\beta=600$, $S=2\times10^{-3}\Omega$ runs 
switching from the mirror to decaying regime and 
from the mirror to firehose regime, shown in 
\figsref{fig:mr_decay}, \ref{fig:mr_to_fh} and \ref{fig:mr_pics}: 
box size $2000\times2000\,d_i^2(0)\approx 81.65\times81.65\,\rho_i^2(0)$, 
cell size $2.6d_i(0)\approx 0.11\rho_i(0)$ ($768^2$ cells), 576 particles per cell.
\end{itemize}
These did not show any significant differences with the lower-resolution/smaller-box 
simulations (including those with the same physical parameters), 
confirming that the latter were adequate. 

\subsection{Effective collision rate}
\label{ap:coll}

The effective collisionalities $\nueff$ due to particle scattering in firehose 
and mirror fluctuations are calculated in the following way. 
We track a number of particles over time intervals $[t,t+\Delta t]$, 
where $\Delta t=100/\Omega(t=0)$, always   
satisfying $\Delta t > \nueff^{-1}$ (confirmed {\em a posteriori}). 
For each particle, we calculate the time $\tau$ over which its first adiabatic invariant 
$\mu=\vperp^2/B$ changes by a factor of 2 (the time evolution of $\mu$ is smoothed 
over time intervals $10/\Omega(0)$, which is of order one Larmor period $2\pi/\Omega(t)$). 
We assemble a histogram of these times and fit it to an exponential $e^{-\tau/\tauc}$. 
We then declare $\nueff=\tauc^{-1}$.

\label{lastpage}

\end{document}